\documentclass[aps,prx,reprint,superscriptaddress]{revtex4-2}
\usepackage{graphicx}
\usepackage{dcolumn}
\usepackage{hyperref}
\usepackage{amsmath}
\usepackage{amssymb}
\usepackage{braket}
\usepackage{upgreek}
\usepackage[version=4]{mhchem}
\usepackage[binary-units]{siunitx}[=v2]
\usepackage{booktabs}
\DeclareSIUnit{\torr}{Torr}
\DeclareSIUnit{\counts}{counts}

\usepackage[utf8]{inputenc}
\usepackage[T1]{fontenc}

\begin{document}

\title{Single-Photon Storage in a Ground-State Vapor Cell Quantum Memory}

\author{Gianni Buser}
\email{gianni.buser@unibas.ch}
\affiliation{Departement Physik, Universit\"{a}t Basel, Klingelbergstr. 82, 4056 Basel, Switzerland.}
\author{Roberto Mottola}
\affiliation{Departement Physik, Universit\"{a}t Basel, Klingelbergstr. 82, 4056 Basel, Switzerland.}
\author{Bj\"{o}rn Cotting}
\affiliation{Departement Physik, Universit\"{a}t Basel, Klingelbergstr. 82, 4056 Basel, Switzerland.}
\author{Janik Wolters}
\affiliation{Departement Physik, Universit\"{a}t Basel, Klingelbergstr. 82, 4056 Basel, Switzerland.}
\affiliation{Deutsches Zentrum f\"{u}r Luft- und Raumfahrt e.V. (DLR), Institute of Optical Sensor Systems, Rutherfordstr. 2, 12489 Berlin, Germany.}
\affiliation{Technische Universit\"{a}t Berlin, Institut f\"{u}r Optik und Atomare Physik, Hardenbergstr. 36, 10623 Berlin, Germany.}
\author{Philipp Treutlein}
\affiliation{Departement Physik, Universit\"{a}t Basel, Klingelbergstr. 82, 4056 Basel, Switzerland.}

\date{\today}

\begin{abstract}
	Interfaced single-photon sources and quantum memories for photons together form a foundational component of quantum technology. Achieving compatibility between heterogeneous, state-of-the-art devices is a long-standing challenge. We built and successfully interfaced a heralded single-photon source based on cavity-enhanced spontaneous parametric down-conversion in ppKTP and a matched memory based on electromagnetically induced transparency in warm \ce{^{87}Rb} vapor. The bandwidth of the photons emitted by the source is \SI{370}{\mega\hertz}, placing its speed in the technologically relevant regime while remaining well within the acceptance bandwidth of the memory. Simultaneously, the experimental complexity is kept low, with all components operating at or above room temperature. Read-out noise of the memory is considerably reduced by exploiting polarization selection rules in the hyperfine structure of spin-polarized atoms. For the first time, we demonstrate single-photon storage and retrieval in a ground-state vapor cell memory, with $g_{c,\,\text{ret}}^{(2)}=\num{0.177\pm0.023}$ demonstrating the single-photon character of the retrieved light. Our platform of single-photon source and atomic memory is attractive for future experiments on room-temperature quantum networks operating at high bandwidth.
\end{abstract}

\maketitle

\section{\label{sec:intro}Introduction}
Quantum memories combined with single photons from high quality sources are versatile and indispensable building blocks across the fields of quantum communication and information. They are at the heart of each node and interconnect in visions of a quantum internet \cite{Kimble2008,Wehner2018}, and central to the standard paradigm of quantum repeaters \cite{Briegel1998,Sangouard2011,Munro2015}. They can be used to synchronize probabilistic gate operations and sources \cite{Nunn2013,Kaneda2017}, and can even improve the indistinguishability of photons emitted by quantum dots through filtering \cite{Gao2019}. Further prospective applications include linear optical quantum computing, metrology, and photon detection \cite{OBrien2007, Bussieres2013}. These applications put different requirements on a memory, calling for a quantitative assessment of memory performance with numerous figures of merit \cite{Simon2010}, including fidelity, efficiency, storage time, bandwidth, and various compatibility parameters. 

Memories implemented in the ground state of room-temperature atomic vapors perform well in terms of fidelity, efficiency, storage times, and bandwidth \cite{Namazi2017,Borregaard2016,Katz2018,Michelberger2015}. Moreover, they are compatible with high quality single-photon sources based on spontaneous parametric downconversion \cite{Michelberger2015,Mottola2020} or semiconductor quantum dots \cite{Akopian2011,Ulrich2014,Jahn2015,Zhai2020}. Together with their technological simplicity, this renders atomic vapor cells a promising memory system for quantum networks, potentially even ones deployed in space \cite{Guendogan2021,Wallnoefer2021}. 
However, a long-standing problem of warm atomic vapor memories is read-out noise arising from four-wave mixing \cite{Phillips2011,Lauk2013} and collisional fluorescence \cite{Rousseau1975,Manz2007}, which degrades the quality of the retrieved photons. Consequently, such memories are commonly tested with laser pulses attenuated to the single-photon level, circumventing the stricter requirements on memory noise imposed by real single photon sources with imperfect efficiencies. Readout noise can be suppressed in cold atom systems \cite{Chaneliere2005,Choi2008,Zhou2012,Wang2019}, or with excited-state storage schemes \cite{Kaczmarek2018,Finkelstein2018}, but these approaches come at the price of much higher experimental complexity, or fundamentally limited storage times, respectively. Demonstrating storage and retrieval of single photons in ground-state vapor cell memories, as characterized by non-classical photon number statistics of the retrieved light, has so far remained elusive \cite{Michelberger2015,Wolters2017,Namazi2017,Thomas2019}.

Here we report the storage and retrieval of single photons in a ground-state atomic vapor cell quantum memory. Our memory scheme suppresses readout noise by exploiting polarization selection rules in the atomic hyperfine structure and by operating at a bandwidth much higher than the excited state's radiative decay rate. We interface the atomic memory with a single photon source based on cavity-enhanced spontaneous parametric downconversion (SPDC), which we built for this purpose with improved operation and performance characteristics compared to our earlier work \cite{Mottola2020}. Single photons from this source are stored in the atomic memory and retrieved with decidedly non-classical photon number statistics, opening up many further possibilities for quantum networking experiments at high bandwidth in a room-temperature system.

\begin{figure}
	\includegraphics[width=\columnwidth]{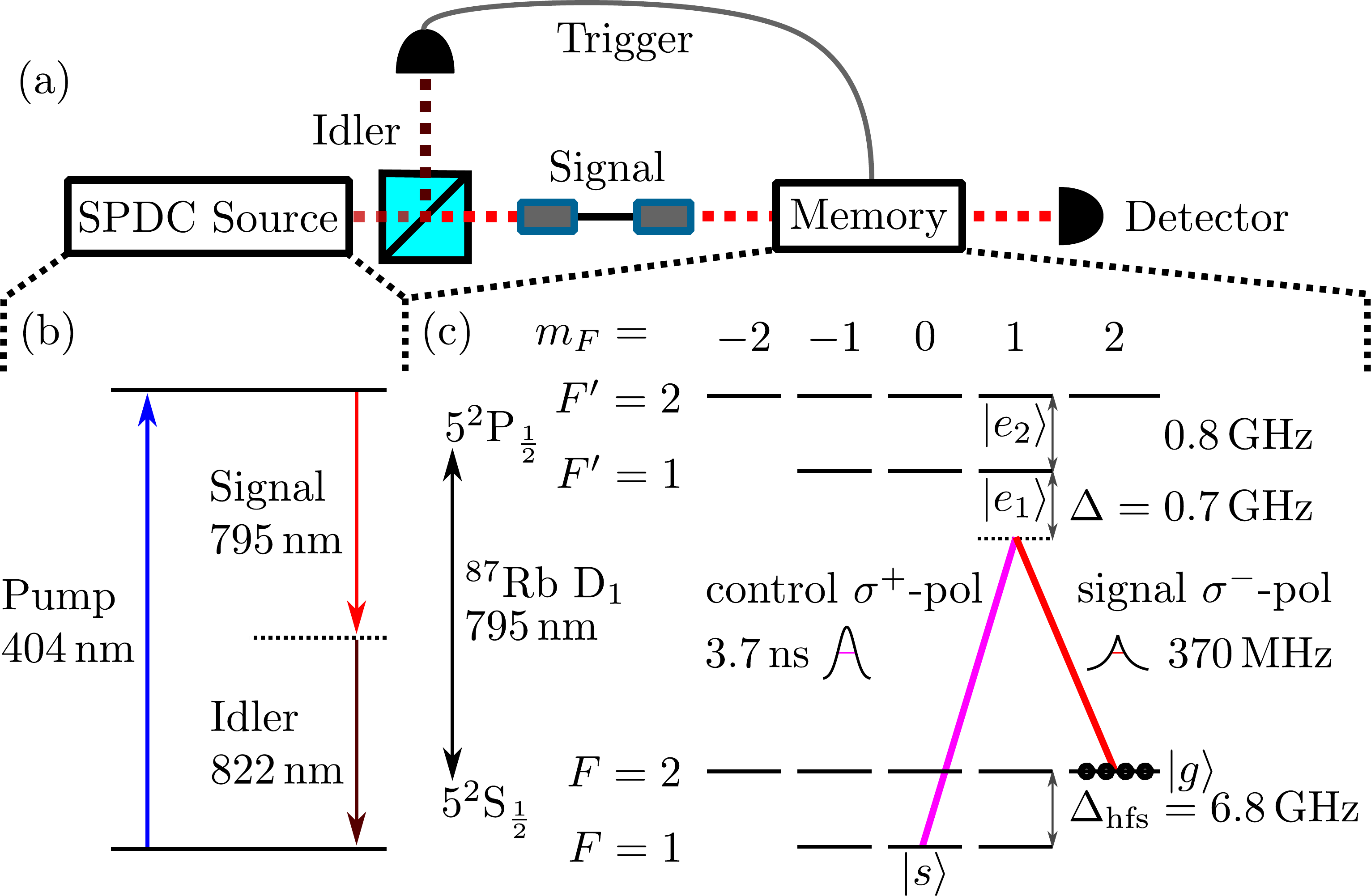}
	\caption{\label{scheme} (a) Basic scheme of the experiment. 
	(b) Energy levels of the photon source. By spontaneous parametric downconversion in ppKTP, \SI{404}{nm} pump photons are converted to \SI{795}{nm} signal and \SI{822}{nm} idler photons. (c) Energy level scheme for the atomic memory. Optical pumping and selection rules are used to isolate a four-level lambda system on the \ce{^{87}Rb} D$_1$ line. The $\sigma^-$ polarized signal is detuned by $\Delta$ from the $F=2\rightarrow F'=1$ transition and the $\sigma^+$ polarized control is equally detuned from $F=1\rightarrow F'=1$. Photons can thus be stored as a spin wave excitation between the initially prepared ground state $\ket{g}$ and the storage state $\ket{s}$ via the excited states $\ket{e_1}$ and $\ket{e_2}$.}
\end{figure}

\section{\label{sec:scheme}Memory Scheme}

An overview of the single-photon source and quantum memory setup is shown in Fig.~\ref{scheme}. 
The memory operates on the \ce{^{87}Rb} D$_1$ line at \SI{795}{\nano\meter} in a hot atomic vapor. We initially prepare the atoms in the stretched Zeeman ground state $\ket{g}=\ket{F=2,m_F=2}$ by optical pumping. This allows us to exploit polarization selection rules to isolate a four-level lambda system formed by the two ground states $\ket{g}$ and $\ket{s}=\ket{F=1,m_F=0}$ and the excited states $\ket{e_1}=\ket{F'=1,m_F'=1}$ and $\ket{e_2}=\ket{F'=2,m_F'=1}$ as shown in Fig.~\ref{scheme} (c). In the storage process, a circularly polarized ($\sigma^-$) signal, that is the single photon to be stored, is reversibly mapped to an atomic ground-state superposition between $\ket{g}$ and $\ket{s}$ by the ($\sigma^+$) control laser. Applying the control laser pulse again after the storage time recreates the photon in the signal mode \cite{Fleischhauer2002}. 

Our scheme overcomes two significant limitations of lambda-scheme atomic memories that do not control the Zeeman state, namely their susceptibility to four-wave mixing noise \cite{Phillips2011,Lauk2013,Michelberger2015} and the presence of parasitic single-photon transitions \cite{Wolters2017}.  The former arises from off-resonant coupling of the strong control laser to the initially prepared atomic state. The latter occur when the signal is absorbed on a transition that would require a selection-rule forbidden mapping to the storage state by the control. In hyperfine lambda-schemes using $\pi$-polarized light this can occur with atoms initially in the $m_F=0$ state \cite{Yan2001}. Both of these problems are addressed simultaneously by controlling the Zeeman state of the atoms and exploiting polarization selection rules \cite{Walther2007}. 

Since the two storage pathways involving $\ket{e_1}$ and $\ket{e_2}$ interfere destructively \cite{Vurgaftman2013}, this approach only works if the detuning of signal and control light is lower or comparable to the excited state splitting, and not in between the states, so that one of the transitions dominates. This interference leads to an effective reduction in the optical depth of the ensemble, but not to absorption without storage. Therefore detunings within a $\si{\giga\hertz}$ range red from $F'=1$ or blue from $F'=2$ can be considered and optimized for signal-to-noise ratio (SNR) and efficiency. The final working point used, $\Delta=-2\pi\times\SI{700}{\mega\hertz}$, is the result of empirical optimization, and the exact value is less crucial than minimizing the two-photon detuning between the signal and control.

We operate our memory in the technologically relevant regime of large bandwidths, typically several hundred MHz, much larger than the excited-state radiative decay rate of the \ce{^{87}Rb} D$_1$ line of $2\pi\times\SI{5.75}{\mega\hertz}$. This allows us to also significantly suppress noise due to collisional fluorescence \cite{Rousseau1975,Manz2007} by time-gating the signal. Overall, our memory scheme thus eliminates several main limitations of previous attempts to store single photons in the ground-state of atomic vapors.

\begin{figure}
	\includegraphics[width=\columnwidth]{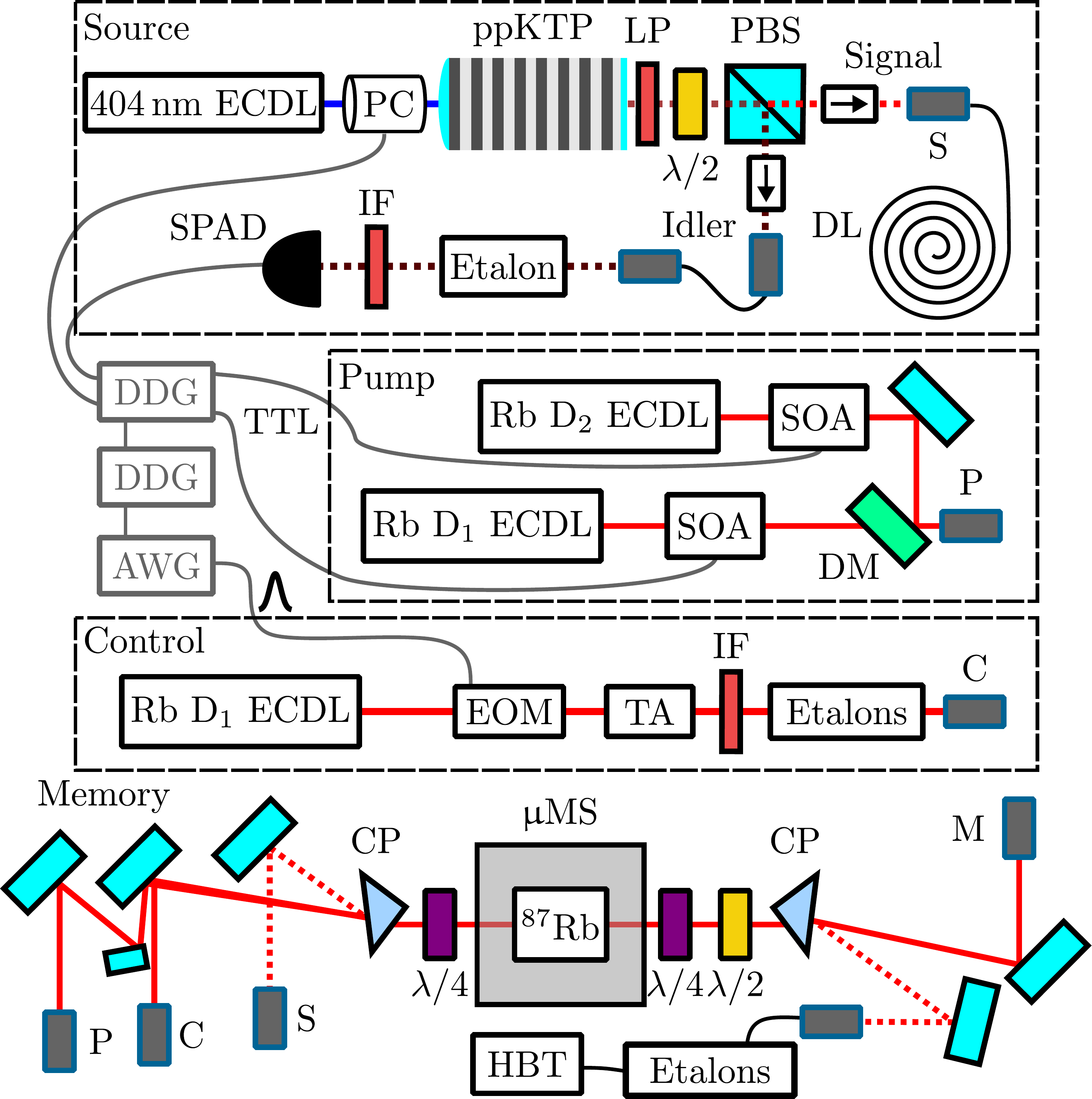}
	\caption{\label{setup} Experimental setup: ECDL external cavity diode laser, PC Pockels cell, ppKTP monolithic periodically poled potassium titanyl phosphate cavity, LP optical longpass, PBS polarizing beamsplitter, IF interference filter, SPAD single photon avalanche diode, DL \SI{60}{\meter} fiber delay line, DDG digital delay generator, AWG arbitrary waveform generator, SOA semiconductor optical amplifier (fiber connections omitted), DM dichroic mirror, EOM electro-optic modulator (fiber connections omitted), TA tapered amplifier, CP calcite polarizing prism, $\lambda/2$ half-wave plate, $\lambda/4$ quarter-wave plate, $\upmu$MS 4-layer mu-metal magnetic shield, HBT Hanbury Brown and Twiss configured single photon detectors, M fiber connection for monitoring the control. The labels S, P, and C represent the fiber connections of the signal, pump, and control to the memory respectively.}
\end{figure}

\section{\label{sec:memsetup}Memory Setup}

The implementation of the memory is further detailed in Fig.~\ref{setup}. The atomic vapor cell at the heart of the memory is a commercial \SI{75}{\milli\meter} long quartz cylinder with \SI{19}{\milli\meter} outer diameter and wedged windows, which contains enriched \ce{^{87}Rb} ($<\SI{1}{\percent}$ \ce{^{85}Rb} specified), \SI{5}{\torr} of \ce{N2} buffer gas, and paraffin coating on the walls. It is housed inside a 4-layer magnetic shield (Twinleaf MS-1L). A simple heater maintains an atomic temperature of \SI{50(1)}{\degreeCelsius}, yielding an optical depth on the signal transition of about 25.
The atomic state preparation is performed with \SI{1}{\milli\meter} $e^{-2}$ diameter, circularly polarized laser beams, pumping on the \ce{^{87}Rb} D$_1$ hyperfine transition $\ket{F=2,m_F}\rightarrow\ket{F'=2,m_F'=m_F+1}$ and repumping on the \ce{^{87}Rb} D$_2$ hyperfine transition $\ket{F=1,m_F}\rightarrow\ket{F',m_F'=m_F+1}$, with around \SI{20}{\milli\watt} and \SI{10}{\milli\watt} CW power, respectively. The effectiveness of the state preparation was characterized at an atomic temperature of \SI{70}{\degreeCelsius} with pump-probe measurements to exclude concerns of radiation trapping at high density \cite{Rosenberry2007}, and is estimated to be $>\SI{98}{\percent}$ initially, decaying exponentially with $\tau=\SI{5(1)}{\micro\second}$. 

The laser setup for pumping and control is schematically shown in Fig.~\ref{setup}.
The pumping beams are generated by CW external cavity diode lasers seeding fiber-integrated semiconductor optical amplifiers (SOAs) for fast switching via the SOA current in response to a herald trigger. They are combined on a dichroic mirror, then fiber coupled and overlapped with the control beam under a small angle of \SI{2.95(15)}{\milli\radian}. The Gaussian control pulses are generated on demand with a fiber-integrated electro-optical amplitude modulator (Jenoptik AM785), controlled by a fast arbitrary pulse generator (PicoQuant PPG512). The FWHM of these pulses measured before the vapor cell is \SI{3.77(4)}{\nano\second}. These pulses are amplified with a tapered amplifier (TA), then spectrally filtered to remove the background of amplified spontaneous emission of the TA with two narrowband interference filters (IFs, \SI{0.5}{\nano\meter} FWHM specified by manufacturer at \SI{795}{\nano\meter}) and two monolithic etalons (\SI{550(10)}{\mega\hertz} bandwidth, \SI{25.5(4)}{\giga\hertz} free spectral range), and finally fiber coupled to bring them to the memory with a maximum possible peak power on the atoms around \SI{680(40)}{\milli\watt}. In the experiments detailed below the power is adjusted to yield a peak Rabi frequency of $\Omega=2\pi\times\SI{400(30)}{\mega\hertz}$ on the $\ket{s}\rightarrow\ket{e_1}$ transition.

Control and signal are initially in orthogonal linear polarizations, and are thus combined on a single calcite prism. Then a quarter-wave plate prepares the required circular polarizations. The signal (control) is focused to a $e^{-2}$ diameter of \SI{480(6)}{\micro\meter} (\SI{520(6)}{\micro\meter}) in the center of the vapor cell by the fiber outcoupling lenses. After the cell further waveplates linearize and align the polarizations, then a second calcite prism separates the signal from the control with a polarization extinction ratio of $>\SI{80}{\deci\bel}$ (characterization limited by detection) after which it is fiber coupled to a spectral filtering stage. These spectral filters consist of three monolithic etalons (\SI{550(10)}{\mega\hertz} bandwidth, \SI{25.5(4)}{\giga\hertz} free spectral range) in series, temperature tuned to the signal frequency. For CW light at the control frequency each of these etalons delivers \SI{-26}{\deci\bel} reduction in intensity. Including the polarization filtering, this results in a total CW control suppression in the signal channel by more than \SI{160}{\deci\bel}. 
Finally, another fiber coupling after the etalons sends the signal to the detection system consisting of two single photon avalanche diodes (SPADs, Excelitas SPCM-AQRH-16) arranged in Hanbury Brown and Twiss configuration. 
Despite the meticulous control filtration, the total transmission of a strong CW probe at the signal frequency from the fiber input of the memory to this output, leaving the atoms warm and unpumped, is $T=\SI{30(3)}{\percent}$.

\section{\label{sec:sourcesetup}SPDC Source}

The single photon source is based on cavity-enhanced SPDC, which provides heralded single photons with high quality, high efficiency, and tunable bandwidth \cite{Slattery2015,Ahlrichs2016,Tsai2018}. The source is an evolution of the one described in \cite{Mottola2020,Cotting2021} with improved reliability and performance. It is carefully tailored to emit signal photons compatible with the atomic memory, at a wavelength of \SI{795}{\nano\meter} fine-tuned to the \ce{^{87}Rb} D$_1$ line and a bandwidth of $\sim$ \SI{370}{MHz} matching the acceptance bandwidth of the memory. 

The heart of the source is a \SI{5}{mm} long periodically poled potassium titanyl phosphate (ppKTP) crystal, polished and coated to form a doubly resonant hemispherical monolithic cavity, see Fig.~\ref{setup}. A periodic poling of \SI{10.1}{\um} is chosen so that the quasi-phase-matching conditions for type-II SPDC are met for signal photons wavelength-matched with the \ce{^{87}Rb} D$_1$ line whereas the bandwidth-matching is given by the cavity linewidth.
In the non-degenerate process \SI{404}{nm} pump photons are down-converted to \SI{795}{nm} (\SI{822}{nm}) signal (idler) photons, illustrated in Fig.~\ref{scheme} (b). 
We pump the crystal with \SI{4.5}{\milli\watt} in a double pass configuration to reach heralding rates of \SI{1.5e5}{\counts\per\second} on average.
The pump frequency is stabilized through a sideband-offset lock on a passively stable reference cavity, allowing for a tunable but locked laser. 
The orthogonally polarized signal and idler photons are split on a polarizing beam splitter cube, see Fig.~\ref{setup}, and coupled each into a polarization maintaining fiber. The herald is filtered with a temperature-stabilized monolithic etalon (\SI{1150(20)}{\mega\hertz} bandwidth, \SI{51(1)}{\giga\hertz} free spectral range) and a narrow-band IF (\SI{0.57(5)}{\nano\meter} FWHM, measured at \SI{822}{\nano\meter}) before being detected with a SPAD. 

We observe that a constant noise floor of uncorrelated photons is emitted by the photon source. In order to suppress this background during photon retrieval a fast Pockels cell acts as switch for the down-conversion process. By rotating the polarization of the pump beam by \ang{90} the down-conversion process is highly suppressed since the phase-matching conditions are not met anymore. It takes about \SI{140}{\nano\second} upon the detection of an idler photon for this switch to turn off the source. As the pump light remains incident on the crystal no thermal drifts are induced by the switching.
Furthermore, optical isolators are placed both in the signal and idler arms. They are used to prevent crosstalk from the lasers preparing the atomic state of the memory and to suppress external cavity modes in the herald path, both of which originate due to spurious back-reflections on the crystal's plane surface.
Even with these additional optical elements the heralding efficiency, that is the probability of having a signal photon exiting the optical fiber to the memory upon the detection of an idler photon, is as high as $\eta_{h} = 40(4)\%$ for a coincidence window of \SI{6.48}{\nano\second}. Such a high $\eta_{h}$ is crucial as even small amounts of memory read-out noise can accumulate when no photon is present most of the time \cite{Jobez2015}. 
We measure the conditional second-order autocorrelation of the signal photons at the memory input to be $g^{(2)}_{\text{input}} = \num{4.21(2)e-2}$, confirming their high quality.

\section{\label{sec:interface}Interfacing source and memory}

Memories for heralded single photons need to react to the detection of the herald, in our case the idler photon from the SPDC source. This places stringent limits on the reaction time of all switching electronics and optics of the memory setup. The time between the detection of an idler photon and the arrival of the signal photon in the vapor cell is about \SI{270}{\nano\second}, as a \SI{60}{\meter} optical fiber is used to route the signal photons from the source to the memory setup. In a storage and retrieval experiment the memory is first initialized by the pumping and repumping lasers for a minimum time of \SI{2}{\micro\second} to ensure that the desired atomic polarization is achieved. During this pumping stage the detection of idler photons is rejected on a hardware level. After this minimum duty cycle of the state preparation, the detection of an idler photon triggers a digital delay generator (DDG, Highland Technology T564, \SI{21}{\nano\second} insertion delay, $<\SI{35}{\pico\second}$ RMS timing jitter) to switch off both the optical pumping and repumping beams as well as the pumping of the source until after the photon is retrieved. All this switching is prioritized to minimize noise, as the extinction ratio of the switches improves over time, and the cables are kept as short as possible. This reactive configuration ensures that the memory remains ready to accept a photon at any time after initialization. The trigger is also relayed to a second DDG for less time critical tasks, including triggering the generation of control pulses and time stamping the idler photon detection with a time-to-digital converter (qutools quTAU, timing resolution \SI{81}{\pico\second}). The typical rate of these experiments is \SI{1.5e5}{\per\second}, set by the chosen heralding rate.

\section{\label{sec:results}Single-photon storage results}

\begin{figure}
	\includegraphics[width=0.9\columnwidth]{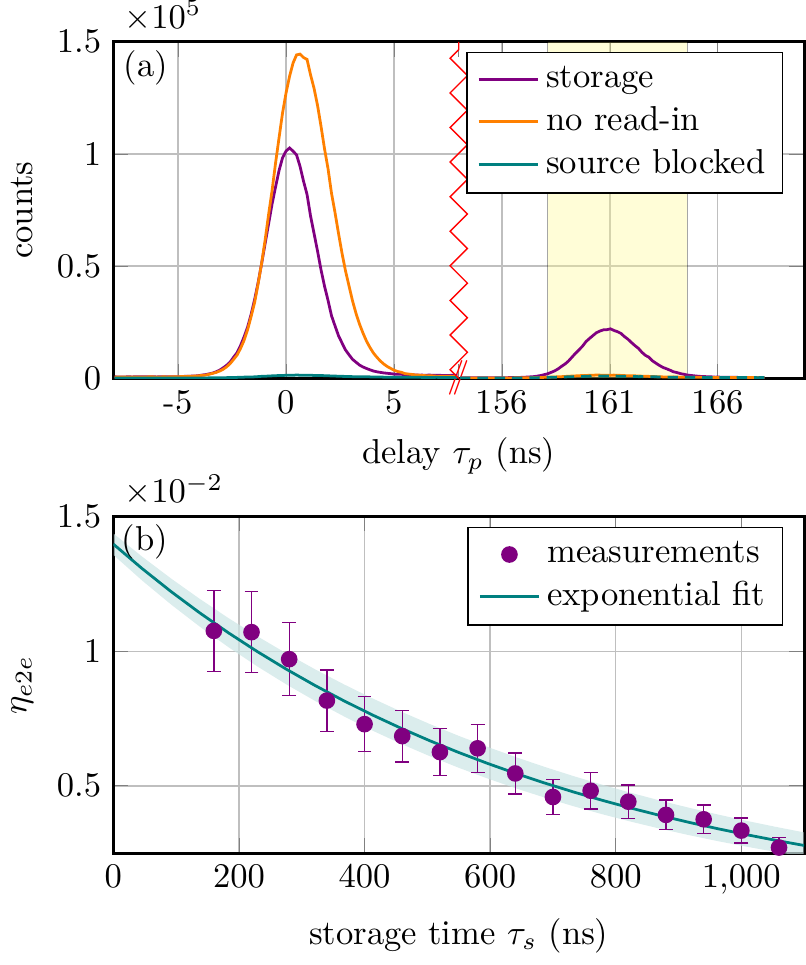}
	\caption{\label{histolin} (a) Photon arrival-time histograms for storage experiments showing the initial leakage of photons through the memory and the retrieval after \SI{160}{\nano\second} of storage. The data are histogrammed in $\SI{162}{\pico\second}$ bins. As no intermediate features are visible on a linear scale the time-axis is broken to show only these peaks. The shaded area marks the $\SI{6.48}{\nano\second}$ ($80\times\SI{81}{\pico\second}$) wide region of interest for the retrieval for which all figures of merit in the text are specified. Almost no noise counts are visible on a linear scale. (b) Memory lifetime as measured by a drop in efficiency and modeled exponentially as $\eta_{e2e}(\tau_s)=\eta_{e2e,0}\exp(-\tau_s/\tau)$. The shaded region marks the $1\sigma$ confidence interval of the fit. Each point represents a storage and retrieval experiment as depicted in (a), where each curve is integrated for \SI{5}{\minute}. These data were collected a day after the main results without optimizing the filters anew, and have thus been scaled to account for a slightly lower setup transmission of about \SI{26}{\percent}.}
\end{figure}

\begin{figure*}
	\includegraphics[width=0.9\textwidth]{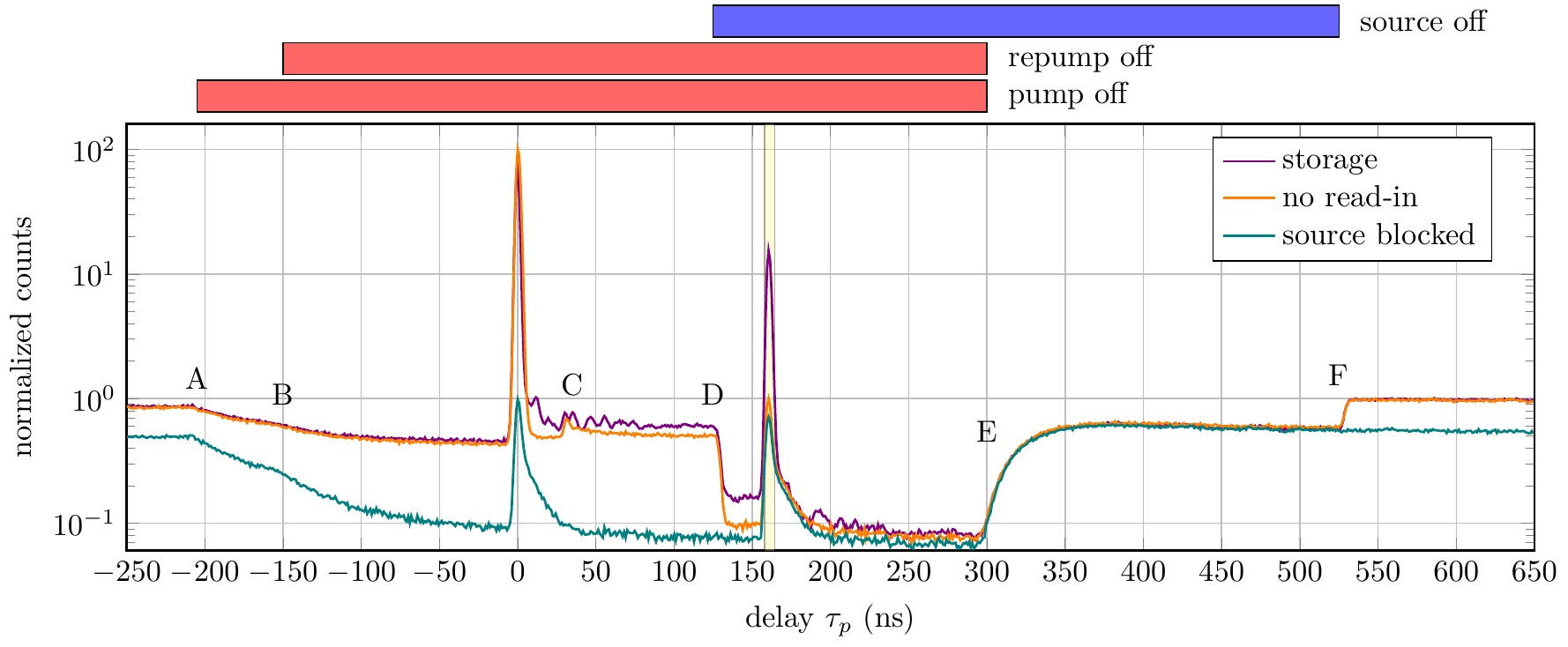}
	\caption{\label{histo} Photon arrival-time histograms for storage experiments on a log-scale. The data are histogrammed in $\SI{1}{\nano\second}$ ($12\times\SI{81}{\pico\second}$) bins, and $\tau_p=0$ represents the time at which an input photon arrives at the detector if it leaks through the setup without being stored or lost. The $y$-axis is normalized to the peak of the no read-in curve in the retrieval window (8325 counts) to yield a proxy for the worst-case signal-to-noise ratio, and is plotted logarithmically so that small features, labeled A-F, can be identified. These illustrate the experimental sequence and are described further in the main text. Colored bars above the plot signify the time windows in terms of detection time wherein the state preparation beams and the photon source are effectively turned off.}
\end{figure*}

Photon arrival-time histograms of the photons detected within \SI{20}{\minute} of integration time are recorded for three scenarios and shown in Figs. \ref{histolin} (a) and \ref{histo}. In the first scenario repeated attempts of storing a heralded photon for $\SI{160}{\nano\second}$ are performed, which we present as a test case for detailed analysis. In a read-out window of \SI{6.48}{\nano\second} (shaded region) and for a total of $N_{\text{herald}}=\num{159752941}$ storage attempts $N_{\text{ret}}=\num{454030}$ photons are retrieved. 

The other two scenarios represent distinct measurements of the noise performance. The source blocked scenario represents the typical readout noise estimation performed for memories, measuring the amount of noise produced when the input is physically blocked. This serves as a comparative measure to other memories, characterizing it in isolation. When the source is blocked the number of memory induced noise counts for the same number of experiments as above is $N_{\text{noise,}\,\text{mem}}=\num{29075}$. The readout noise floor is then $\mu_{\text{mem}}=N_{\text{noise,}\,\text{mem}}/N_{\text{herald}}=\num{1.82(18)e-4}$. This however does not capture noise stemming either from the source itself in the form of uncorrelated photons at the signal frequency, nor from spurious back reflections of light originating from the memory setup in the source. 

The total noise is estimated in a measurement omitting the read-in control pulse, labeled as the no read-in scenario. This induces a small systematic error as the atomic response to the read-out pulse is also influenced by the read-in pulse. The no read-in curve therefore delivers a slight overestimation of the total amount of noise present in the read-out window. The number of noise counts detected in this curve's read-out window is $N_{\text{noise,}\,\text{tot}}=\num{38634}$ yielding a total noise floor of $\mu_{\text{tot}}=\num{2.42(24)e-4}$. With this we calculate the end-to-end efficiency defined as
\begin{equation}
	\eta_{\text{e2e}}=\frac{N_{\text{ret}}-N_{\text{noise,}\,\text{tot}}}{N_{\text{herald}}\eta_h\eta_{\text{det}}},
\end{equation}
to be $\eta_{e2e}=\SI{1.1(2)}{\percent}$. Here $\eta_{\text{det}}$ is the quantum efficiency of the single photon detectors specified as \SI{60\pm 6}{\percent}, which is the dominant source of uncertainty. The signal-to-noise ratio of the combined experiment is $\text{SNR}=\frac{N_{\text{ret}}-N_{\text{noise,tot}}}{N_{\text{noise,tot}}}=\num{10.8\pm1.5}$ which bodes well for the quality of the retrieved photons. Indeed, the conditioned autocorrelation of the retrieved photons is $g_{c,\,\text{ret}}^{(2)}=\num{0.177(23)}$ confirming that the memory emission is dominated by single photons. This is the first experimental confirmation of the viability of retrieving single photons from ground-state memories in hot atomic vapor.

A measurement of the exponential $1/e$ lifetime of the memory is shown in Fig. \ref{histolin} (b). The fit yields $\tau=\SI{680(50)}{\nano\second}$ lifetime and an initial efficiency of $\eta_{e2e,0}=\SI{1.4(4)}{\percent}$. This lifetime is close to what is expected for a memory limited by atomic motion out of the interaction region, but may be slightly shortened by the lifetime of the initial ground state preparation which has not been exhaustively optimized for longevity. Accounting for the technical losses in the setup by dividing out the measured transmission and extrapolating to zero storage yields a total internal efficiency of $\eta_{\text{int}}=\eta_{e2e,0}/T=\SI{4.7(14)}{\percent}$. The conditioned autocorrelation of the retrieved photons eventually increases for long storage times. For $\tau_s=\SI{280}{\nano\second}$ it is still low at $g_{c,\,\text{ret}}^{(2)}=\num{0.171\pm0.028}$, but by $\tau_s=\SI{700}{\nano\second}$ it increases to $g_{c,\,\text{ret}}^{(2)}=\num{0.503(93)}$. The lifetime could therefore also be seen as roughly the time for which $g_{c,\,\text{ret}}^{(2)} \leq 0.5$.

The photon number statistics of the retrieved light depend on the photon number statistics of the signal pulse, as well as those of the noise and potentially the process at its origin too. An exact and physically simple model for the case of an incoherent admixture of noise is derived in \cite{Michelberger2015}. It treats the noise as being generated independently from the storage and retrieval processes, the additional light being added to the read-out by straightforward superposition. It reads
\onecolumngrid
\begin{equation}
	\label{eq:g2model}
	g_{c,\,\text{ret,}\,\text{theo}}^{(2)}=\frac{\left(N_{\text{ret}}-N_{\text{noise}}\right)^2g^{(2)}_{\text{input}}+2N_{\text{noise}}\left(N_{\text{ret}}-N_{\text{noise}}\right)+N_{\text{noise}}^2g^{(2)}_{\text{noise}}}{N_{\text{ret}}^2}.
\end{equation}
\twocolumngrid
Here, $g^{(2)}_{\text{input}}$ has been established during the characterization of the source; the statistics of the noise $g_{c,\,\text{noise}}^{(2)}$, however, are not measured directly, as insufficient noise counts accumulate within a reasonable integration time to evaluate them meaningfully. Sources of noise for which we would expect $g_{c,\,\text{noise}}^{(2)}=1$ are limited to leaked control laser light. Known possible thermal noise sources include uncorrelated SPDC photons from the source, as well as collisional fluorescence and four-wave mixing induced by the control. We distinguish these scenarios on the memory side by scanning the final filter etalon with the SPDC source blocked. We observe collisional fluorescence peaks on resonance with the natural oscillators $F'\rightarrow F=2$ \cite{Rousseau1975} which seem to constitute the entirety of the noise at the signal frequency. Scanning the etalon to the control frequency, we also observe the control laser and collisional fluorescence on $F'\rightarrow F=1$, but these are well suppressed when the final etalon is set to the signal frequency. No further peaks are visible after the filters, in particular confirming that four-wave mixing is well suppressed by the memory scheme. We therefore expect $g_{c,\,\text{noise}}^{(2)}=2$ as non-thermal noise sources are well excluded. Moreover, if the noise can be modeled incoherently this is a conservative assumption as it represents the worst case. This approach yields an expected value of $g_{c,\,\text{ret,}\,\text{theo}}^{(2)}=\num{0.204\pm0.029}$, which is in excellent agreement with our observed statistics. In the limits $g^{(2)}_{\text{input}}\rightarrow 0$, $g_{c,\,\text{noise}}^{(2)}\rightarrow 2$ the exact model reduces to $g_{c,\,\text{ret,}\,\text{theo}}^{(2)}\approx2/(\text{SNR}+1)$, illustrating the link between the statistical measure and the memory performance directly visible in the data. Note that were the memory limited by a coherent noise source such as four-wave mixing gain, incoherent models would drastically underestimate the $g^{(2)}$ of the retrieved light \cite{Michelberger2015}.

Figure~\ref{histo} makes small features in the data visible yielding additional insights into the main features. The peaks at $\tau_p=0$ and $\tau_p=\SI{161}{\nano\second}$, also displayed in Fig.~\ref{histolin} (a), are now revealed to be followed by exponential tails. These counts correspond to fluorescence from the atomic excited state. Due to the high bandwidth, the majority of this noise is avoided thanks to its temporal separation from the signal, highlighting an innate advantage of high-speed memories. Further labeled features correspond to steps in the experimental sequence and technical effects. In the region A (B) the herald photon has already triggered the electronics, and we see exponentially decaying fluorescence from the atoms as the pumping (repumping) beam is switched off. In the region C, after the photon is stored, the discrepancy between the storage and no read-in curves corresponds to unintentional read-out of the stored photon. The end-to-end (internal) efficiency of this unintentional read-out is $\SI{0.38(5)}{\percent}$ $(\SI{1.26(17)}{\percent})$. It is caused by ringing in the EOM switching the control and its limited extinction ratio. At longer storage times the end-to-end efficiency of this read-out saturates at about \SI{1}{\percent}. The feature present in both curves right beneath the label C at around $\tau_p=\SI{30}{\nano\second}$ is due to afterpulsing of the SPADs. The steep feature labeled D corresponds to the Pockels cell switching off the SPDC source. The remaining discrepancy between the no-read and source blocked curves estimates the secondary effect of the first control pulse on the atoms. The rising feature labeled E corresponds to both pumping lasers being switched back on after the retrieval is complete. Finally, the feature labeled F marks the source switching back on.

\section{\label{sec:discussion}Discussion and Simulation} 

To analyze limitations of our experiment and moreover guide future development we simulate the storage and retrieval process numerically. We consider an atomic four-level system following the description in \cite{Rakher2013}, also taking into account the transverse profile of both light fields, similar to \cite{Nunn2008}. We include both collisional broadening induced by the buffer gas and inhomogeneous Doppler-broadening of the vapor. The latter is taken into account by introducing different velocity classes of the atoms, as shown in \cite{Gorshkov2007a}. We focus on the scenario of forward retrieval.
The storage and the retrieval processes are much faster than the mean free time between collisions of about \SI{20}{\nano\second} in the vapor cell used for this experiment. This allows us to assume that the individual atoms do not change their velocities during these processes. However, during the storage time of \SI{160}{\nano\second} we assume that the atoms rethermalize fully. 
The numerical simulation is implemented by solving the temporal derivatives appearing in the equations of motion with a partially implicit second-order Runge-Kutta method. For the spatial derivatives spectral collocation is used, allowing us to replace them with Chebyshev differentiation matrices as described in \cite{Nunn2008}.

\begin{figure}
	\includegraphics[width=\columnwidth]{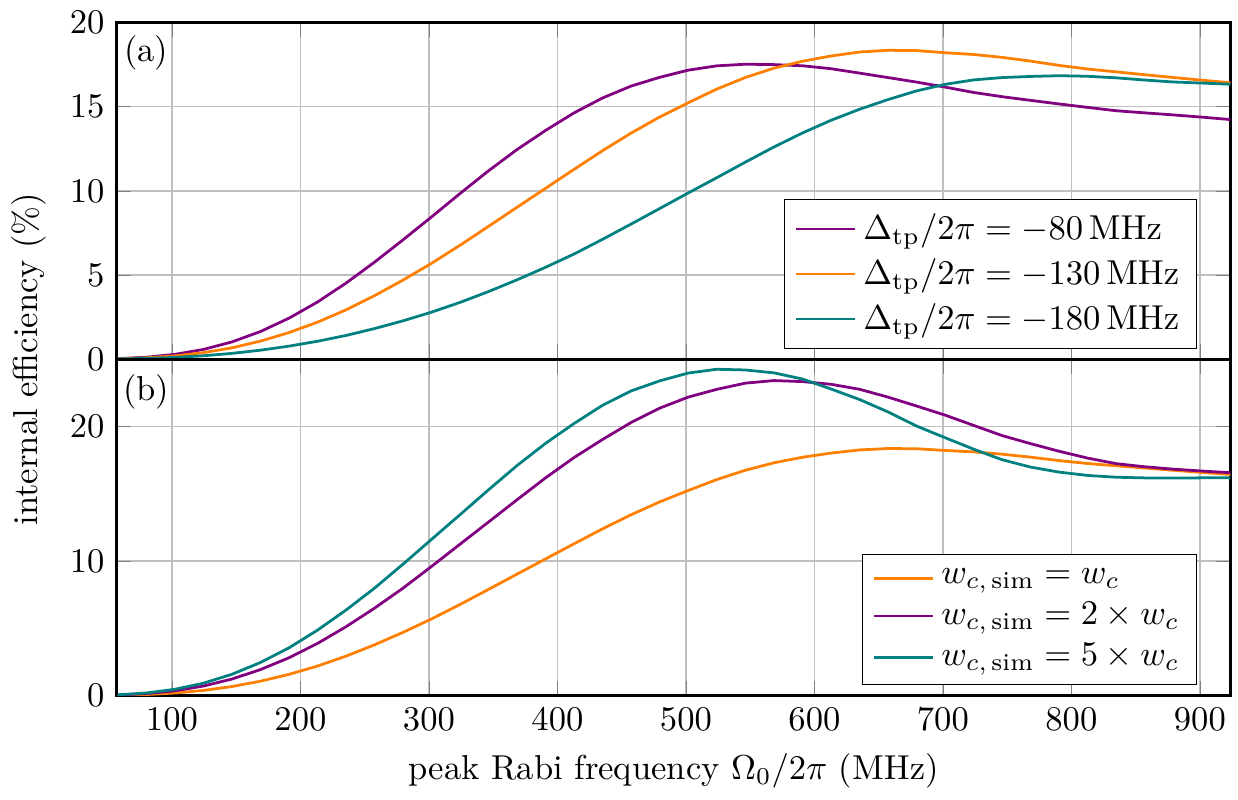}
	\caption{\label{simulation} (a) Simulation of the internal memory efficiency using the known experimental parameters, as a function of the peak control Rabi frequency for different two-photon detunings. (b) Simulation of the internal memory efficiency as above, but for different control beam waists and setting $\Delta_{\text{tp}}/2\pi=\SI{-130}{\mega\hertz}$, where $w_c=\SI{260(3)}{\micro\meter}$ is the experimental waist of the control in the vapor cell center.}
\end{figure}

We use the simulation to evaluate the total efficiency as a function of the peak control Rabi frequency. Herein the temporal width of the Gaussian control pulse is fixed to \SI{3.77}{\nano\second}, and the time alignment between signal and control pulse is optimized for each simulation point. Seeking plausible routes to improvements, other experimentally controllable parameters are varied. We find that the two-photon detuning has a strong impact on the attainable performance, as can be seen in Fig.~\ref{simulation} (a). Minimization of the two-photon detuning intuitively corresponds to bringing the memory interaction into resonance, note however, that following the convention of \cite{Gorshkov2007b} we define this parameter as the difference between the control and signal detunings on their respective single-photon transitions, $\Delta_{\text{tp}}=\Delta_c-\Delta_s$. In the regime of our experiment where $\Omega\approx\Delta<\Delta_{\text{hfs}}$, the presence of the control pulse induces a significant, time-dependent level-shift, which by this definition appears as a shift of the optimal detuning from $\Delta_{\text{tp}}=0$. Furthermore, different control beam waists are considered. The simulation confirms that by choosing a larger control beam waist the total efficiency can be improved. Fig.~\ref{simulation} (b) shows the performance for different control waist for optimal two-photon detuning. With a control beam significantly larger than the signal, more atoms experience the optimal control Rabi frequency, leading to a more efficient process. 

For the parameters of the experiment shown in Figs. \ref{histolin} (a) and \ref{histo} the measured value of the internal efficiency falls short of that predicted by our model. This leads us to identify effects not captured by the simulation. One is the unintentional read-out of the signal during the storage time by the leaking control light which is present due to a combination of ringing in the EOM as well as its finite extinction ratio (about \SI{25}{\deci\bel}). A second issue is the quality of the temporal alignment of the control pulse to the signal, mainly limited by the $>\SI{350}{\pico\second}$ jitter of the single photon detector measuring the idler photon. Based on tests with laser pulses where this jitter source can be eliminated, we expect this amount of variance in the temporal alignment to decrease the efficiency by a factor of $<0.90(5)$. These issues, alongside the difficulty of stating an experimental value of $\Delta_{\text{tp}}$ with any certainty, makes modeling the exact experimental situation difficult. Experimentally, the signal frequency is set once, whereupon the control frequency is optimized on the storage efficiency at fixed $\Omega$ to compensate for drifts, without the possibility of a direct measurement of the two-photon detuning. Our simulations will guide future improvements by identifying the regimes were the best performance is expected, accounting for the interdependence of the many optimization parameters. This analysis highlights three main areas where technical improvements can still be implemented: lasers, optical switches, and detectors. A more powerful control laser would enable us to define a larger and more homogeneous interaction region. Optical switches without ringing and with a better on/off ratio would protect the spinwave during storage. Faster single photon detectors would improve the time alignment of the control pulse to the single signal photon. In particular, the currently remaining jitter on a control pulse, removing the jitter of the idler detection, is measured to be only \SI{126(3)}{\pico\second}. The state-of-the-art in timing resolution of single photon detection is more than an order of magnitude smaller. 

Upon addressing technical matters, a further improvement of the efficiency is possible by increasing the optical depth. Technically this is straightforward via the atomic temperature, but is bound to affect other figures of merit as well, in particular the readout noise. Due to their interdependence, parameters such as the detuning from the excited states and Rabi frequency would again require optimization, and accurate performance predictions concerning noise, fidelity, and lifetime predicate dedicated models not currently implemented. Additionally, some gains could be expected from shaping the temporal profile of the control pulses. This is fraught with ambitious demands on the time resolution of the pulse generation. A more straightforward change, still requiring some modifications to the electronic systems though, would be to optimize the amplitude of the read-in and read-out pulses independently. Although the optimal processes are well known to be linked by time-reversal symmetry both theoretically \cite{Gorshkov2008} and experimentally \cite{Phillips2008}, the matter is less cut-and-dried in the experimentally implemented high-bandwidth case with realizable Gaussian control pulses, where a full pulse shape optimization is technically prohibitive and a practical distinction between programmed and effective pulses is warranted.

\section{\label{sec:outlook}Summary and Outlook}
In summary, we have for the first time demonstrated the interfacing of a heralded single photon source and a technologically simple atomic quantum memory with ground state storage successfully maintaining the single photon nature of the input upon retrieval. Through simulations we have laid a road map for future improvements that will simultaneously realize state-of-the-art efficiency. Moreover, the fidelity of the memory will be determined more directly with Hong-Ou-Mandel experiments \cite{Hong1987}. Modern vapor cell fabrication techniques can produce cells with diameters $<\SI{1}{\milli\meter}$ \cite{Moeller2017}. Even with the currently available control power, which determines the mode size yielding the optimal control intensity, this would match the cell size to the optical mode. Under such conditions the atoms would remain in the interaction region despite their motion during storage. With corresponding improvements to the longevity of the Zeeman state preparation, which is known to be short but not currently limiting, a memory lifetime on the order of milliseconds is well within reach judging by similar systems \cite{Katz2018}. Moreover, Zeeman-state preserving anti-relaxation coatings on cells have shown far longer relaxation times \cite{Corsini2013}. Possible applications of our source-memory system thus include photon synchronization and use in high-bandwidth quantum networks for quantum information processing and simulation. The multi-photon rate enhancement provided by memories is given by the time-bandwidth product $B$, scaled by the memory efficiency \cite{Nunn2013}. Due to the high bandwidth operation of our memory we achieve $B=\num{250(20)}$ despite the non-optimized storage time, which would be sufficient to demonstrate enhanced synchronized rates directly.

\begin{acknowledgments}

We thank C. M\"{u}ller, T. Kroh, A. Ahlrichs, S. Ramelow, and O. Benson for initiating us into the mysteries of SPDC sources. We thank R. Warburton, L. Zhai, G. Nguyen, and C. Spinnler for fruitful discussions and acknowledge financial support from NCCR QSIT. 

\end{acknowledgments}


\begin{thebibliography}{53}
	\makeatletter
	\providecommand \@ifxundefined [1]{%
		\@ifx{#1\undefined}
	}%
	\providecommand \@ifnum [1]{%
		\ifnum #1\expandafter \@firstoftwo
		\else \expandafter \@secondoftwo
		\fi
	}%
	\providecommand \@ifx [1]{%
		\ifx #1\expandafter \@firstoftwo
		\else \expandafter \@secondoftwo
		\fi
	}%
	\providecommand \natexlab [1]{#1}%
	\providecommand \enquote  [1]{``#1''}%
	\providecommand \bibnamefont  [1]{#1}%
	\providecommand \bibfnamefont [1]{#1}%
	\providecommand \citenamefont [1]{#1}%
	\providecommand \href@noop [0]{\@secondoftwo}%
	\providecommand \href [0]{\begingroup \@sanitize@url \@href}%
	\providecommand \@href[1]{\@@startlink{#1}\@@href}%
	\providecommand \@@href[1]{\endgroup#1\@@endlink}%
	\providecommand \@sanitize@url [0]{\catcode `\\12\catcode `\$12\catcode
		`\&12\catcode `\#12\catcode `\^12\catcode `\_12\catcode `\%12\relax}%
	\providecommand \@@startlink[1]{}%
	\providecommand \@@endlink[0]{}%
	\providecommand \url  [0]{\begingroup\@sanitize@url \@url }%
	\providecommand \@url [1]{\endgroup\@href {#1}{\urlprefix }}%
	\providecommand \urlprefix  [0]{URL }%
	\providecommand \Eprint [0]{\href }%
	\providecommand \doibase [0]{https://doi.org/}%
	\providecommand \selectlanguage [0]{\@gobble}%
	\providecommand \bibinfo  [0]{\@secondoftwo}%
	\providecommand \bibfield  [0]{\@secondoftwo}%
	\providecommand \translation [1]{[#1]}%
	\providecommand \BibitemOpen [0]{}%
	\providecommand \bibitemStop [0]{}%
	\providecommand \bibitemNoStop [0]{.\EOS\space}%
	\providecommand \EOS [0]{\spacefactor3000\relax}%
	\providecommand \BibitemShut  [1]{\csname bibitem#1\endcsname}%
	\let\auto@bib@innerbib\@empty

\bibitem [{\citenamefont {Kimble}(2008)}]{Kimble2008}%
\BibitemOpen
\bibfield  {author} {\bibinfo {author} {\bibfnamefont {H.~J.}\ \bibnamefont
		{Kimble}},\ }\bibfield  {title} {\bibinfo {title} {The quantum internet},\
}\href {https://doi.org/10.1038/nature07127} {\bibfield  {journal} {\bibinfo
		{journal} {Nature}\ }\textbf {\bibinfo {volume} {453}},\ \bibinfo {pages}
	{1023} (\bibinfo {year} {2008})}\BibitemShut {NoStop}%
\bibitem [{\citenamefont {Wehner}\ \emph {et~al.}(2018)\citenamefont {Wehner},
	\citenamefont {Elkouss},\ and\ \citenamefont {Hanson}}]{Wehner2018}%
\BibitemOpen
\bibfield  {author} {\bibinfo {author} {\bibfnamefont {S.}~\bibnamefont
		{Wehner}}, \bibinfo {author} {\bibfnamefont {D.}~\bibnamefont {Elkouss}},\
	and\ \bibinfo {author} {\bibfnamefont {R.}~\bibnamefont {Hanson}},\
}\bibfield  {title} {\bibinfo {title} {Quantum internet: A vision for the
		road ahead},\ }\href {https://doi.org/10.1126/science.aam9288} {\bibfield
	{journal} {\bibinfo  {journal} {Science}\ }\textbf {\bibinfo {volume}
		{362}},\ \bibinfo {pages} {eaam9288} (\bibinfo {year} {2018})}\BibitemShut
{NoStop}%
\bibitem [{\citenamefont {Briegel}\ \emph {et~al.}(1998)\citenamefont
	{Briegel}, \citenamefont {D\"{u}r}, \citenamefont {Cirac},\ and\
	\citenamefont {Zoller}}]{Briegel1998}%
\BibitemOpen
\bibfield  {author} {\bibinfo {author} {\bibfnamefont {H.-J.}\ \bibnamefont
		{Briegel}}, \bibinfo {author} {\bibfnamefont {W.}~\bibnamefont {D\"{u}r}},
	\bibinfo {author} {\bibfnamefont {J.~I.}\ \bibnamefont {Cirac}},\ and\
	\bibinfo {author} {\bibfnamefont {P.}~\bibnamefont {Zoller}},\ }\bibfield
{title} {\bibinfo {title} {Quantum repeaters: {T}he role of imperfect local
		operations in quantum communication},\ }\href
{https://doi.org/10.1103/PhysRevLett.81.5932} {\bibfield  {journal} {\bibinfo
		{journal} {Physical Review Letters}\ }\textbf {\bibinfo {volume} {81}},\
	\bibinfo {pages} {5932} (\bibinfo {year} {1998})}\BibitemShut {NoStop}%
\bibitem [{\citenamefont {Sangouard}\ \emph {et~al.}(2011)\citenamefont
	{Sangouard}, \citenamefont {Simon}, \citenamefont {de~Riedmatten},\ and\
	\citenamefont {Gisin}}]{Sangouard2011}%
\BibitemOpen
\bibfield  {author} {\bibinfo {author} {\bibfnamefont {N.}~\bibnamefont
		{Sangouard}}, \bibinfo {author} {\bibfnamefont {C.}~\bibnamefont {Simon}},
	\bibinfo {author} {\bibfnamefont {H.}~\bibnamefont {de~Riedmatten}},\ and\
	\bibinfo {author} {\bibfnamefont {N.}~\bibnamefont {Gisin}},\ }\bibfield
{title} {\bibinfo {title} {Quantum repeaters based on atomic ensembles and
		linear optics},\ }\href {https://doi.org/10.1103/RevModPhys.83.33} {\bibfield
	{journal} {\bibinfo  {journal} {Reviews of Modern Physics}\ }\textbf
	{\bibinfo {volume} {83}},\ \bibinfo {pages} {33} (\bibinfo {year}
	{2011})}\BibitemShut {NoStop}%
\bibitem [{\citenamefont {Munro}\ \emph {et~al.}(2015)\citenamefont {Munro},
	\citenamefont {Azuma}, \citenamefont {Tamaki},\ and\ \citenamefont
	{Nemoto}}]{Munro2015}%
\BibitemOpen
\bibfield  {author} {\bibinfo {author} {\bibfnamefont {W.~J.}\ \bibnamefont
		{Munro}}, \bibinfo {author} {\bibfnamefont {K.}~\bibnamefont {Azuma}},
	\bibinfo {author} {\bibfnamefont {K.}~\bibnamefont {Tamaki}},\ and\ \bibinfo
	{author} {\bibfnamefont {K.}~\bibnamefont {Nemoto}},\ }\bibfield  {title}
{\bibinfo {title} {Inside quantum repeaters},\ }\href
{https://doi.org/10.1109/jstqe.2015.2392076} {\bibfield  {journal} {\bibinfo
		{journal} {IEEE Journal of Selected Topics in Quantum Electronics}\ }\textbf
	{\bibinfo {volume} {21}},\ \bibinfo {pages} {78} (\bibinfo {year}
	{2015})}\BibitemShut {NoStop}%
\bibitem [{\citenamefont {Nunn}\ \emph {et~al.}(2013)\citenamefont {Nunn},
	\citenamefont {Langford}, \citenamefont {Kolthammer}, \citenamefont
	{Champion}, \citenamefont {Sprague}, \citenamefont {Michelberger},
	\citenamefont {Jin}, \citenamefont {England},\ and\ \citenamefont
	{Walmsley}}]{Nunn2013}%
\BibitemOpen
\bibfield  {author} {\bibinfo {author} {\bibfnamefont {J.}~\bibnamefont
		{Nunn}}, \bibinfo {author} {\bibfnamefont {N.~K.}\ \bibnamefont {Langford}},
	\bibinfo {author} {\bibfnamefont {W.~S.}\ \bibnamefont {Kolthammer}},
	\bibinfo {author} {\bibfnamefont {T.~F.~M.}\ \bibnamefont {Champion}},
	\bibinfo {author} {\bibfnamefont {M.~R.}\ \bibnamefont {Sprague}}, \bibinfo
	{author} {\bibfnamefont {P.~S.}\ \bibnamefont {Michelberger}}, \bibinfo
	{author} {\bibfnamefont {X.-M.}\ \bibnamefont {Jin}}, \bibinfo {author}
	{\bibfnamefont {D.~G.}\ \bibnamefont {England}},\ and\ \bibinfo {author}
	{\bibfnamefont {I.~A.}\ \bibnamefont {Walmsley}},\ }\bibfield  {title}
{\bibinfo {title} {Enhancing multiphoton rates with quantum memories},\
}\href {https://doi.org/10.1103/PhysRevLett.110.133601} {\bibfield  {journal}
	{\bibinfo  {journal} {Physical Review Letters}\ }\textbf {\bibinfo {volume}
		{110}},\ \bibinfo {pages} {133601} (\bibinfo {year} {2013})}\BibitemShut
{NoStop}%
\bibitem [{\citenamefont {Kaneda}\ \emph {et~al.}(2017)\citenamefont {Kaneda},
	\citenamefont {Xu}, \citenamefont {Chapman},\ and\ \citenamefont
	{Kwiat}}]{Kaneda2017}%
\BibitemOpen
\bibfield  {author} {\bibinfo {author} {\bibfnamefont {F.}~\bibnamefont
		{Kaneda}}, \bibinfo {author} {\bibfnamefont {F.}~\bibnamefont {Xu}}, \bibinfo
	{author} {\bibfnamefont {J.}~\bibnamefont {Chapman}},\ and\ \bibinfo {author}
	{\bibfnamefont {P.~G.}\ \bibnamefont {Kwiat}},\ }\bibfield  {title} {\bibinfo
	{title} {Quantum-memory-assisted multi-photon generation for efficient
		quantum information processing},\ }\href
{https://doi.org/10.1364/optica.4.001034} {\bibfield  {journal} {\bibinfo
		{journal} {Optica}\ }\textbf {\bibinfo {volume} {4}},\ \bibinfo {pages}
	{1034} (\bibinfo {year} {2017})}\BibitemShut {NoStop}%
\bibitem [{\citenamefont {Gao}\ \emph {et~al.}(2019)\citenamefont {Gao},
	\citenamefont {Lazo-Arjona}, \citenamefont {Brecht}, \citenamefont
	{Kaczmarek}, \citenamefont {Thomas}, \citenamefont {Nunn}, \citenamefont
	{Ledingham}, \citenamefont {Saunders},\ and\ \citenamefont
	{Walmsley}}]{Gao2019}%
\BibitemOpen
\bibfield  {author} {\bibinfo {author} {\bibfnamefont {S.}~\bibnamefont
		{Gao}}, \bibinfo {author} {\bibfnamefont {O.}~\bibnamefont {Lazo-Arjona}},
	\bibinfo {author} {\bibfnamefont {B.}~\bibnamefont {Brecht}}, \bibinfo
	{author} {\bibfnamefont {K.~T.}\ \bibnamefont {Kaczmarek}}, \bibinfo {author}
	{\bibfnamefont {S.~E.}\ \bibnamefont {Thomas}}, \bibinfo {author}
	{\bibfnamefont {J.}~\bibnamefont {Nunn}}, \bibinfo {author} {\bibfnamefont
		{P.~M.}\ \bibnamefont {Ledingham}}, \bibinfo {author} {\bibfnamefont {D.~J.}\
		\bibnamefont {Saunders}},\ and\ \bibinfo {author} {\bibfnamefont {I.~A.}\
		\bibnamefont {Walmsley}},\ }\bibfield  {title} {\bibinfo {title} {Optimal
		coherent filtering for single noisy photons},\ }\href
{https://doi.org/10.1103/PhysRevLett.123.213604} {\bibfield  {journal}
	{\bibinfo  {journal} {Physical Review Letters}\ }\textbf {\bibinfo {volume}
		{123}},\ \bibinfo {pages} {213604} (\bibinfo {year} {2019})}\BibitemShut
{NoStop}%
\bibitem [{\citenamefont {O'Brien}(2007)}]{OBrien2007}%
\BibitemOpen
\bibfield  {author} {\bibinfo {author} {\bibfnamefont {J.~L.}\ \bibnamefont
		{O'Brien}},\ }\bibfield  {title} {\bibinfo {title} {Optical quantum
		computing},\ }\href {https://doi.org/10.1126/science.1142892} {\bibfield
	{journal} {\bibinfo  {journal} {Science}\ }\textbf {\bibinfo {volume}
		{318}},\ \bibinfo {pages} {1567} (\bibinfo {year} {2007})}\BibitemShut
{NoStop}%
\bibitem [{\citenamefont {Bussi{\`{e}}res}\ \emph {et~al.}(2013)\citenamefont
	{Bussi{\`{e}}res}, \citenamefont {Sangouard}, \citenamefont {Afzelius},
	\citenamefont {de~Riedmatten}, \citenamefont {Simon},\ and\ \citenamefont
	{Tittel}}]{Bussieres2013}%
\BibitemOpen
\bibfield  {author} {\bibinfo {author} {\bibfnamefont {F.}~\bibnamefont
		{Bussi{\`{e}}res}}, \bibinfo {author} {\bibfnamefont {N.}~\bibnamefont
		{Sangouard}}, \bibinfo {author} {\bibfnamefont {M.}~\bibnamefont {Afzelius}},
	\bibinfo {author} {\bibfnamefont {H.}~\bibnamefont {de~Riedmatten}}, \bibinfo
	{author} {\bibfnamefont {C.}~\bibnamefont {Simon}},\ and\ \bibinfo {author}
	{\bibfnamefont {W.}~\bibnamefont {Tittel}},\ }\bibfield  {title} {\bibinfo
	{title} {Prospective applications of optical quantum memories},\ }\href
{https://doi.org/10.1080/09500340.2013.856482} {\bibfield  {journal}
	{\bibinfo  {journal} {Journal of Modern Optics}\ }\textbf {\bibinfo {volume}
		{60}},\ \bibinfo {pages} {1519} (\bibinfo {year} {2013})}\BibitemShut
{NoStop}%
\bibitem [{\citenamefont {Simon}\ \emph {et~al.}(2010)\citenamefont {Simon},
	\citenamefont {Afzelius}, \citenamefont {Appel}, \citenamefont {de~la
		Giroday}, \citenamefont {Dewhurst}, \citenamefont {Gisin}, \citenamefont
	{Hu}, \citenamefont {Jelezko}, \citenamefont {Kr\"{o}ll}, \citenamefont
	{M\"{u}ller}, \citenamefont {Nunn}, \citenamefont {Polzik}, \citenamefont
	{Rarity}, \citenamefont {Riedmatten}, \citenamefont {Rosenfeld},
	\citenamefont {Shields}, \citenamefont {Sk\"{o}ld}, \citenamefont
	{Stevenson}, \citenamefont {Thew}, \citenamefont {Walmsley}, \citenamefont
	{Weber}, \citenamefont {Weinfurter}, \citenamefont {Wrachtrup},\ and\
	\citenamefont {Young}}]{Simon2010}%
\BibitemOpen
\bibfield  {author} {\bibinfo {author} {\bibfnamefont {C.}~\bibnamefont
		{Simon}}, \bibinfo {author} {\bibfnamefont {M.}~\bibnamefont {Afzelius}},
	\bibinfo {author} {\bibfnamefont {J.}~\bibnamefont {Appel}}, \bibinfo
	{author} {\bibfnamefont {A.~B.}\ \bibnamefont {de~la Giroday}}, \bibinfo
	{author} {\bibfnamefont {S.~J.}\ \bibnamefont {Dewhurst}}, \bibinfo {author}
	{\bibfnamefont {N.}~\bibnamefont {Gisin}}, \bibinfo {author} {\bibfnamefont
		{C.~Y.}\ \bibnamefont {Hu}}, \bibinfo {author} {\bibfnamefont
		{F.}~\bibnamefont {Jelezko}}, \bibinfo {author} {\bibfnamefont
		{S.}~\bibnamefont {Kr\"{o}ll}}, \bibinfo {author} {\bibfnamefont {J.~H.}\
		\bibnamefont {M\"{u}ller}}, \bibinfo {author} {\bibfnamefont
		{J.}~\bibnamefont {Nunn}}, \bibinfo {author} {\bibfnamefont {E.~S.}\
		\bibnamefont {Polzik}}, \bibinfo {author} {\bibfnamefont {J.~G.}\
		\bibnamefont {Rarity}}, \bibinfo {author} {\bibfnamefont {H.~D.}\
		\bibnamefont {Riedmatten}}, \bibinfo {author} {\bibfnamefont
		{W.}~\bibnamefont {Rosenfeld}}, \bibinfo {author} {\bibfnamefont {A.~J.}\
		\bibnamefont {Shields}}, \bibinfo {author} {\bibfnamefont {N.}~\bibnamefont
		{Sk\"{o}ld}}, \bibinfo {author} {\bibfnamefont {R.~M.}\ \bibnamefont
		{Stevenson}}, \bibinfo {author} {\bibfnamefont {R.}~\bibnamefont {Thew}},
	\bibinfo {author} {\bibfnamefont {I.~A.}\ \bibnamefont {Walmsley}}, \bibinfo
	{author} {\bibfnamefont {M.~C.}\ \bibnamefont {Weber}}, \bibinfo {author}
	{\bibfnamefont {H.}~\bibnamefont {Weinfurter}}, \bibinfo {author}
	{\bibfnamefont {J.}~\bibnamefont {Wrachtrup}},\ and\ \bibinfo {author}
	{\bibfnamefont {R.~J.}\ \bibnamefont {Young}},\ }\bibfield  {title} {\bibinfo
	{title} {Quantum memories},\ }\href
{https://doi.org/10.1140/epjd/e2010-00103-y} {\bibfield  {journal} {\bibinfo
		{journal} {The European Physical Journal D}\ }\textbf {\bibinfo {volume}
		{58}},\ \bibinfo {pages} {1} (\bibinfo {year} {2010})}\BibitemShut {NoStop}%
\bibitem [{\citenamefont {Namazi}\ \emph {et~al.}(2017)\citenamefont {Namazi},
	\citenamefont {Kupchak}, \citenamefont {Jordaan}, \citenamefont
	{Shahrokhshahi},\ and\ \citenamefont {Figueroa}}]{Namazi2017}%
\BibitemOpen
\bibfield  {author} {\bibinfo {author} {\bibfnamefont {M.}~\bibnamefont
		{Namazi}}, \bibinfo {author} {\bibfnamefont {C.}~\bibnamefont {Kupchak}},
	\bibinfo {author} {\bibfnamefont {B.}~\bibnamefont {Jordaan}}, \bibinfo
	{author} {\bibfnamefont {R.}~\bibnamefont {Shahrokhshahi}},\ and\ \bibinfo
	{author} {\bibfnamefont {E.}~\bibnamefont {Figueroa}},\ }\bibfield  {title}
{\bibinfo {title} {Ultralow-noise room-temperature quantum memory for
		polarization qubits},\ }\href
{https://doi.org/10.1103/physrevapplied.8.034023} {\bibfield  {journal}
	{\bibinfo  {journal} {Physical Review Applied}\ }\textbf {\bibinfo {volume}
		{8}},\ \bibinfo {pages} {034023} (\bibinfo {year} {2017})}\BibitemShut
{NoStop}%
\bibitem [{\citenamefont {Borregaard}\ \emph {et~al.}(2016)\citenamefont
	{Borregaard}, \citenamefont {Zugenmaier}, \citenamefont {Petersen},
	\citenamefont {Shen}, \citenamefont {Vasilakis}, \citenamefont {Jensen},
	\citenamefont {Polzik},\ and\ \citenamefont {S{\o}rensen}}]{Borregaard2016}%
\BibitemOpen
\bibfield  {author} {\bibinfo {author} {\bibfnamefont {J.}~\bibnamefont
		{Borregaard}}, \bibinfo {author} {\bibfnamefont {M.}~\bibnamefont
		{Zugenmaier}}, \bibinfo {author} {\bibfnamefont {J.~M.}\ \bibnamefont
		{Petersen}}, \bibinfo {author} {\bibfnamefont {H.}~\bibnamefont {Shen}},
	\bibinfo {author} {\bibfnamefont {G.}~\bibnamefont {Vasilakis}}, \bibinfo
	{author} {\bibfnamefont {K.}~\bibnamefont {Jensen}}, \bibinfo {author}
	{\bibfnamefont {E.~S.}\ \bibnamefont {Polzik}},\ and\ \bibinfo {author}
	{\bibfnamefont {A.~S.}\ \bibnamefont {S{\o}rensen}},\ }\bibfield  {title}
{\bibinfo {title} {Scalable photonic network architecture based on motional
		averaging in room temperature gas},\ }\href
{https://doi.org/10.1038/ncomms11356} {\bibfield  {journal} {\bibinfo
		{journal} {Nature Communications}\ }\textbf {\bibinfo {volume} {7}},\
	\bibinfo {pages} {11356} (\bibinfo {year} {2016})}\BibitemShut {NoStop}%
\bibitem [{\citenamefont {Katz}\ and\ \citenamefont
	{Firstenberg}(2018)}]{Katz2018}%
\BibitemOpen
\bibfield  {author} {\bibinfo {author} {\bibfnamefont {O.}~\bibnamefont
		{Katz}}\ and\ \bibinfo {author} {\bibfnamefont {O.}~\bibnamefont
		{Firstenberg}},\ }\bibfield  {title} {\bibinfo {title} {Light storage for one
		second in room-temperature alkali vapor},\ }\href
{https://doi.org/10.1038/s41467-018-04458-4} {\bibfield  {journal} {\bibinfo
		{journal} {Nature Communications}\ }\textbf {\bibinfo {volume} {9}},\
	\bibinfo {pages} {2074} (\bibinfo {year} {2018})}\BibitemShut {NoStop}%
\bibitem [{\citenamefont {Michelberger}\ \emph {et~al.}(2015)\citenamefont
	{Michelberger}, \citenamefont {Champion}, \citenamefont {Sprague},
	\citenamefont {Kaczmarek}, \citenamefont {Barbieri}, \citenamefont {Jin},
	\citenamefont {England}, \citenamefont {Kolthammer}, \citenamefont
	{Saunders}, \citenamefont {Nunn},\ and\ \citenamefont
	{Walmsley}}]{Michelberger2015}%
\BibitemOpen
\bibfield  {author} {\bibinfo {author} {\bibfnamefont {P.~S.}\ \bibnamefont
		{Michelberger}}, \bibinfo {author} {\bibfnamefont {T.~F.~M.}\ \bibnamefont
		{Champion}}, \bibinfo {author} {\bibfnamefont {M.~R.}\ \bibnamefont
		{Sprague}}, \bibinfo {author} {\bibfnamefont {K.~T.}\ \bibnamefont
		{Kaczmarek}}, \bibinfo {author} {\bibfnamefont {M.}~\bibnamefont {Barbieri}},
	\bibinfo {author} {\bibfnamefont {X.~M.}\ \bibnamefont {Jin}}, \bibinfo
	{author} {\bibfnamefont {D.~G.}\ \bibnamefont {England}}, \bibinfo {author}
	{\bibfnamefont {W.~S.}\ \bibnamefont {Kolthammer}}, \bibinfo {author}
	{\bibfnamefont {D.~J.}\ \bibnamefont {Saunders}}, \bibinfo {author}
	{\bibfnamefont {J.}~\bibnamefont {Nunn}},\ and\ \bibinfo {author}
	{\bibfnamefont {I.~A.}\ \bibnamefont {Walmsley}},\ }\bibfield  {title}
{\bibinfo {title} {Interfacing {GH}z-bandwidth heralded single photons with a
		warm vapour {R}aman memory},\ }\href
{https://doi.org/10.1088/1367-2630/17/4/043006} {\bibfield  {journal}
	{\bibinfo  {journal} {New Journal of Physics}\ }\textbf {\bibinfo {volume}
		{17}},\ \bibinfo {pages} {043006} (\bibinfo {year} {2015})}\BibitemShut
{NoStop}%
\bibitem [{\citenamefont {Mottola}\ \emph {et~al.}(2020)\citenamefont
	{Mottola}, \citenamefont {Buser}, \citenamefont {M\"{u}ller}, \citenamefont
	{Kroh}, \citenamefont {Ahlrichs}, \citenamefont {Ramelow}, \citenamefont
	{Benson}, \citenamefont {Treutlein},\ and\ \citenamefont
	{Wolters}}]{Mottola2020}%
\BibitemOpen
\bibfield  {author} {\bibinfo {author} {\bibfnamefont {R.}~\bibnamefont
		{Mottola}}, \bibinfo {author} {\bibfnamefont {G.}~\bibnamefont {Buser}},
	\bibinfo {author} {\bibfnamefont {C.}~\bibnamefont {M\"{u}ller}}, \bibinfo
	{author} {\bibfnamefont {T.}~\bibnamefont {Kroh}}, \bibinfo {author}
	{\bibfnamefont {A.}~\bibnamefont {Ahlrichs}}, \bibinfo {author}
	{\bibfnamefont {S.}~\bibnamefont {Ramelow}}, \bibinfo {author} {\bibfnamefont
		{O.}~\bibnamefont {Benson}}, \bibinfo {author} {\bibfnamefont
		{P.}~\bibnamefont {Treutlein}},\ and\ \bibinfo {author} {\bibfnamefont
		{J.}~\bibnamefont {Wolters}},\ }\bibfield  {title} {\bibinfo {title} {An
		efficient, tunable, and robust source of narrow-band photon pairs at the
		$^{87}${R}b {D}1 line},\ }\href {https://doi.org/10.1364/oe.384081}
{\bibfield  {journal} {\bibinfo  {journal} {Optics Express}\ }\textbf
	{\bibinfo {volume} {28}},\ \bibinfo {pages} {3159} (\bibinfo {year}
	{2020})}\BibitemShut {NoStop}%
\bibitem [{\citenamefont {Akopian}\ \emph {et~al.}(2011)\citenamefont
	{Akopian}, \citenamefont {Wang}, \citenamefont {Rastelli}, \citenamefont
	{Schmidt},\ and\ \citenamefont {Zwiller}}]{Akopian2011}%
\BibitemOpen
\bibfield  {author} {\bibinfo {author} {\bibfnamefont {N.}~\bibnamefont
		{Akopian}}, \bibinfo {author} {\bibfnamefont {L.}~\bibnamefont {Wang}},
	\bibinfo {author} {\bibfnamefont {A.}~\bibnamefont {Rastelli}}, \bibinfo
	{author} {\bibfnamefont {O.~G.}\ \bibnamefont {Schmidt}},\ and\ \bibinfo
	{author} {\bibfnamefont {V.}~\bibnamefont {Zwiller}},\ }\bibfield  {title}
{\bibinfo {title} {Hybrid semiconductor-atomic interface: Slowing down single
		photons from a quantum dot},\ }\href
{https://doi.org/10.1038/nphoton.2011.16} {\bibfield  {journal} {\bibinfo
		{journal} {Nature Photonics}\ }\textbf {\bibinfo {volume} {5}},\ \bibinfo
	{pages} {230} (\bibinfo {year} {2011})}\BibitemShut {NoStop}%
\bibitem [{\citenamefont {Ulrich}\ \emph {et~al.}(2014)\citenamefont {Ulrich},
	\citenamefont {Weiler}, \citenamefont {Oster}, \citenamefont {Jetter},
	\citenamefont {Urvoy}, \citenamefont {L\"{o}w},\ and\ \citenamefont
	{Michler}}]{Ulrich2014}%
\BibitemOpen
\bibfield  {author} {\bibinfo {author} {\bibfnamefont {S.~M.}\ \bibnamefont
		{Ulrich}}, \bibinfo {author} {\bibfnamefont {S.}~\bibnamefont {Weiler}},
	\bibinfo {author} {\bibfnamefont {M.}~\bibnamefont {Oster}}, \bibinfo
	{author} {\bibfnamefont {M.}~\bibnamefont {Jetter}}, \bibinfo {author}
	{\bibfnamefont {A.}~\bibnamefont {Urvoy}}, \bibinfo {author} {\bibfnamefont
		{R.}~\bibnamefont {L\"{o}w}},\ and\ \bibinfo {author} {\bibfnamefont
		{P.}~\bibnamefont {Michler}},\ }\bibfield  {title} {\bibinfo {title}
	{Spectroscopy of the {D}$_1$ transition of cesium by dressed-state resonance
		fluorescence from a single ({I}n,{G}a){A}s/{G}a{A}s quantum dot},\ }\href
{https://doi.org/10.1103/physrevb.90.125310} {\bibfield  {journal} {\bibinfo
		{journal} {Physical Review B}\ }\textbf {\bibinfo {volume} {90}},\ \bibinfo
	{pages} {125310} (\bibinfo {year} {2014})}\BibitemShut {NoStop}%
\bibitem [{\citenamefont {Jahn}\ \emph {et~al.}(2015)\citenamefont {Jahn},
	\citenamefont {Munsch}, \citenamefont {B{\'{e}}guin}, \citenamefont
	{Kuhlmann}, \citenamefont {Renggli}, \citenamefont {Huo}, \citenamefont
	{Ding}, \citenamefont {Trotta}, \citenamefont {Reindl}, \citenamefont
	{Schmidt}, \citenamefont {Rastelli}, \citenamefont {Treutlein},\ and\
	\citenamefont {Warburton}}]{Jahn2015}%
\BibitemOpen
\bibfield  {author} {\bibinfo {author} {\bibfnamefont {J.-P.}\ \bibnamefont
		{Jahn}}, \bibinfo {author} {\bibfnamefont {M.}~\bibnamefont {Munsch}},
	\bibinfo {author} {\bibfnamefont {L.}~\bibnamefont {B{\'{e}}guin}}, \bibinfo
	{author} {\bibfnamefont {A.~V.}\ \bibnamefont {Kuhlmann}}, \bibinfo {author}
	{\bibfnamefont {M.}~\bibnamefont {Renggli}}, \bibinfo {author} {\bibfnamefont
		{Y.}~\bibnamefont {Huo}}, \bibinfo {author} {\bibfnamefont {F.}~\bibnamefont
		{Ding}}, \bibinfo {author} {\bibfnamefont {R.}~\bibnamefont {Trotta}},
	\bibinfo {author} {\bibfnamefont {M.}~\bibnamefont {Reindl}}, \bibinfo
	{author} {\bibfnamefont {O.~G.}\ \bibnamefont {Schmidt}}, \bibinfo {author}
	{\bibfnamefont {A.}~\bibnamefont {Rastelli}}, \bibinfo {author}
	{\bibfnamefont {P.}~\bibnamefont {Treutlein}},\ and\ \bibinfo {author}
	{\bibfnamefont {R.~J.}\ \bibnamefont {Warburton}},\ }\bibfield  {title}
{\bibinfo {title} {An artificial {R}b atom in a semiconductor with
		lifetime-limited linewidth},\ }\href
{https://doi.org/10.1103/physrevb.92.245439} {\bibfield  {journal} {\bibinfo
		{journal} {Physical Review B}\ }\textbf {\bibinfo {volume} {92}},\ \bibinfo
	{pages} {245439} (\bibinfo {year} {2015})}\BibitemShut {NoStop}%
\bibitem [{\citenamefont {Zhai}\ \emph {et~al.}(2020)\citenamefont {Zhai},
	\citenamefont {L\"{o}bl}, \citenamefont {Jahn}, \citenamefont {Huo},
	\citenamefont {Treutlein}, \citenamefont {Schmidt}, \citenamefont
	{Rastelli},\ and\ \citenamefont {Warburton}}]{Zhai2020}%
\BibitemOpen
\bibfield  {author} {\bibinfo {author} {\bibfnamefont {L.}~\bibnamefont
		{Zhai}}, \bibinfo {author} {\bibfnamefont {M.~C.}\ \bibnamefont {L\"{o}bl}},
	\bibinfo {author} {\bibfnamefont {J.-P.}\ \bibnamefont {Jahn}}, \bibinfo
	{author} {\bibfnamefont {Y.}~\bibnamefont {Huo}}, \bibinfo {author}
	{\bibfnamefont {P.}~\bibnamefont {Treutlein}}, \bibinfo {author}
	{\bibfnamefont {O.~G.}\ \bibnamefont {Schmidt}}, \bibinfo {author}
	{\bibfnamefont {A.}~\bibnamefont {Rastelli}},\ and\ \bibinfo {author}
	{\bibfnamefont {R.~J.}\ \bibnamefont {Warburton}},\ }\bibfield  {title}
{\bibinfo {title} {Large-range frequency tuning of a narrow-linewidth quantum
		emitter},\ }\href {https://doi.org/10.1063/5.0017995} {\bibfield  {journal}
	{\bibinfo  {journal} {Applied Physics Letters}\ }\textbf {\bibinfo {volume}
		{117}},\ \bibinfo {pages} {083106} (\bibinfo {year} {2020})}\BibitemShut
{NoStop}%
\bibitem [{\citenamefont {G{\"{u}}ndo{\u{g}}an}\ \emph
	{et~al.}(2021)\citenamefont {G{\"{u}}ndo{\u{g}}an}, \citenamefont {Sidhu},
	\citenamefont {Henderson}, \citenamefont {Mazzarella}, \citenamefont
	{Wolters}, \citenamefont {Oi},\ and\ \citenamefont
	{Krutzik}}]{Guendogan2021}%
\BibitemOpen
\bibfield  {author} {\bibinfo {author} {\bibfnamefont {M.}~\bibnamefont
		{G{\"{u}}ndo{\u{g}}an}}, \bibinfo {author} {\bibfnamefont {J.~S.}\
		\bibnamefont {Sidhu}}, \bibinfo {author} {\bibfnamefont {V.}~\bibnamefont
		{Henderson}}, \bibinfo {author} {\bibfnamefont {L.}~\bibnamefont
		{Mazzarella}}, \bibinfo {author} {\bibfnamefont {J.}~\bibnamefont {Wolters}},
	\bibinfo {author} {\bibfnamefont {D.~K.~L.}\ \bibnamefont {Oi}},\ and\
	\bibinfo {author} {\bibfnamefont {M.}~\bibnamefont {Krutzik}},\ }\bibfield
{title} {\bibinfo {title} {Proposal for space-borne quantum memories for
		global quantum networking},\ }\bibfield  {journal} {\bibinfo  {journal} {npj
		Quantum Information}\ }\textbf {\bibinfo {volume} {7}},\ \href
{https://doi.org/10.1038/s41534-021-00460-9} {10.1038/s41534-021-00460-9}
(\bibinfo {year} {2021})\BibitemShut {NoStop}%
\bibitem [{\citenamefont {Wallnöfer}\ \emph {et~al.}(2021)\citenamefont
	{Wallnöfer}, \citenamefont {Hahn}, \citenamefont {Gündoğan}, \citenamefont
	{Sidhu}, \citenamefont {Krüger}, \citenamefont {Walk}, \citenamefont
	{Eisert},\ and\ \citenamefont {Wolters}}]{Wallnoefer2021}%
\BibitemOpen
\bibfield  {author} {\bibinfo {author} {\bibfnamefont {J.}~\bibnamefont
		{Wallnöfer}}, \bibinfo {author} {\bibfnamefont {F.}~\bibnamefont {Hahn}},
	\bibinfo {author} {\bibfnamefont {M.}~\bibnamefont {Gündoğan}}, \bibinfo
	{author} {\bibfnamefont {J.~S.}\ \bibnamefont {Sidhu}}, \bibinfo {author}
	{\bibfnamefont {F.}~\bibnamefont {Krüger}}, \bibinfo {author} {\bibfnamefont
		{N.}~\bibnamefont {Walk}}, \bibinfo {author} {\bibfnamefont {J.}~\bibnamefont
		{Eisert}},\ and\ \bibinfo {author} {\bibfnamefont {J.}~\bibnamefont
		{Wolters}},\ }\bibfield  {title} {\bibinfo {title} {Simulating quantum
		repeater strategies for multiple satellites},\ }\href@noop {} {\  (\bibinfo
	{year} {2021})},\ \Eprint {https://arxiv.org/abs/2110.15806}
{arXiv:2110.15806 [quant-ph]} \BibitemShut {NoStop}%
\bibitem [{\citenamefont {Phillips}\ \emph {et~al.}(2011)\citenamefont
	{Phillips}, \citenamefont {Gorshkov},\ and\ \citenamefont
	{Novikova}}]{Phillips2011}%
\BibitemOpen
\bibfield  {author} {\bibinfo {author} {\bibfnamefont {N.~B.}\ \bibnamefont
		{Phillips}}, \bibinfo {author} {\bibfnamefont {A.~V.}\ \bibnamefont
		{Gorshkov}},\ and\ \bibinfo {author} {\bibfnamefont {I.}~\bibnamefont
		{Novikova}},\ }\bibfield  {title} {\bibinfo {title} {Light storage in an
		optically thick atomic ensemble under conditions of electromagnetically
		induced transparency and four-wave mixing},\ }\href
{https://doi.org/10.1103/physreva.83.063823} {\bibfield  {journal} {\bibinfo
		{journal} {Physical Review A}\ }\textbf {\bibinfo {volume} {83}},\ \bibinfo
	{pages} {063823} (\bibinfo {year} {2011})}\BibitemShut {NoStop}%
\bibitem [{\citenamefont {Lauk}\ \emph {et~al.}(2013)\citenamefont {Lauk},
	\citenamefont {O'Brien},\ and\ \citenamefont {Fleischhauer}}]{Lauk2013}%
\BibitemOpen
\bibfield  {author} {\bibinfo {author} {\bibfnamefont {N.}~\bibnamefont
		{Lauk}}, \bibinfo {author} {\bibfnamefont {C.}~\bibnamefont {O'Brien}},\ and\
	\bibinfo {author} {\bibfnamefont {M.}~\bibnamefont {Fleischhauer}},\
}\bibfield  {title} {\bibinfo {title} {Fidelity of photon propagation in
		electromagnetically induced transparency in the presence of four-wave
		mixing},\ }\href {https://doi.org/10.1103/physreva.88.013823} {\bibfield
	{journal} {\bibinfo  {journal} {Physical Review A}\ }\textbf {\bibinfo
		{volume} {88}},\ \bibinfo {pages} {013823} (\bibinfo {year}
	{2013})}\BibitemShut {NoStop}%
\bibitem [{\citenamefont {Rousseau}\ \emph {et~al.}(1975)\citenamefont
	{Rousseau}, \citenamefont {Patterson},\ and\ \citenamefont
	{Williams}}]{Rousseau1975}%
\BibitemOpen
\bibfield  {author} {\bibinfo {author} {\bibfnamefont {D.~L.}\ \bibnamefont
		{Rousseau}}, \bibinfo {author} {\bibfnamefont {G.~D.}\ \bibnamefont
		{Patterson}},\ and\ \bibinfo {author} {\bibfnamefont {P.~F.}\ \bibnamefont
		{Williams}},\ }\bibfield  {title} {\bibinfo {title} {Resonance {R}aman
		scattering and collision-induced redistribution scattering in {I}$_2$},\
}\href {https://doi.org/10.1103/physrevlett.34.1306} {\bibfield  {journal}
	{\bibinfo  {journal} {Physical Review Letters}\ }\textbf {\bibinfo {volume}
		{34}},\ \bibinfo {pages} {1306} (\bibinfo {year} {1975})}\BibitemShut
{NoStop}%
\bibitem [{\citenamefont {Manz}\ \emph {et~al.}(2007)\citenamefont {Manz},
	\citenamefont {Fernholz}, \citenamefont {Schmiedmayer},\ and\ \citenamefont
	{Pan}}]{Manz2007}%
\BibitemOpen
\bibfield  {author} {\bibinfo {author} {\bibfnamefont {S.}~\bibnamefont
		{Manz}}, \bibinfo {author} {\bibfnamefont {T.}~\bibnamefont {Fernholz}},
	\bibinfo {author} {\bibfnamefont {J.}~\bibnamefont {Schmiedmayer}},\ and\
	\bibinfo {author} {\bibfnamefont {J.-W.}\ \bibnamefont {Pan}},\ }\bibfield
{title} {\bibinfo {title} {Collisional decoherence during writing and reading
		quantum states},\ }\href {https://doi.org/10.1103/physreva.75.040101}
{\bibfield  {journal} {\bibinfo  {journal} {Physical Review A}\ }\textbf
	{\bibinfo {volume} {75}},\ \bibinfo {pages} {040101(R)} (\bibinfo {year}
	{2007})}\BibitemShut {NoStop}%
\bibitem [{\citenamefont {Chaneli{\`{e}}re}\ \emph {et~al.}(2005)\citenamefont
	{Chaneli{\`{e}}re}, \citenamefont {Matsukevich}, \citenamefont {Jenkins},
	\citenamefont {Lan}, \citenamefont {Kennedy},\ and\ \citenamefont
	{Kuzmich}}]{Chaneliere2005}%
\BibitemOpen
\bibfield  {author} {\bibinfo {author} {\bibfnamefont {T.}~\bibnamefont
		{Chaneli{\`{e}}re}}, \bibinfo {author} {\bibfnamefont {D.~N.}\ \bibnamefont
		{Matsukevich}}, \bibinfo {author} {\bibfnamefont {S.~D.}\ \bibnamefont
		{Jenkins}}, \bibinfo {author} {\bibfnamefont {S.-Y.}\ \bibnamefont {Lan}},
	\bibinfo {author} {\bibfnamefont {T.~A.~B.}\ \bibnamefont {Kennedy}},\ and\
	\bibinfo {author} {\bibfnamefont {A.}~\bibnamefont {Kuzmich}},\ }\bibfield
{title} {\bibinfo {title} {Storage and retrieval of single photons
		transmitted between remote quantum memories},\ }\href
{https://doi.org/10.1038/nature04315} {\bibfield  {journal} {\bibinfo
		{journal} {Nature}\ }\textbf {\bibinfo {volume} {438}},\ \bibinfo {pages}
	{833} (\bibinfo {year} {2005})}\BibitemShut {NoStop}%
\bibitem [{\citenamefont {Choi}\ \emph {et~al.}(2008)\citenamefont {Choi},
	\citenamefont {Deng}, \citenamefont {Laurat},\ and\ \citenamefont
	{Kimble}}]{Choi2008}%
\BibitemOpen
\bibfield  {author} {\bibinfo {author} {\bibfnamefont {K.~S.}\ \bibnamefont
		{Choi}}, \bibinfo {author} {\bibfnamefont {H.}~\bibnamefont {Deng}}, \bibinfo
	{author} {\bibfnamefont {J.}~\bibnamefont {Laurat}},\ and\ \bibinfo {author}
	{\bibfnamefont {H.~J.}\ \bibnamefont {Kimble}},\ }\bibfield  {title}
{\bibinfo {title} {Mapping photonic entanglement into and out of a quantum
		memory},\ }\href {https://doi.org/10.1038/nature06670} {\bibfield  {journal}
	{\bibinfo  {journal} {Nature}\ }\textbf {\bibinfo {volume} {452}},\ \bibinfo
	{pages} {67} (\bibinfo {year} {2008})}\BibitemShut {NoStop}%
\bibitem [{\citenamefont {Zhou}\ \emph {et~al.}(2012)\citenamefont {Zhou},
	\citenamefont {Zhang}, \citenamefont {Liu}, \citenamefont {Chen},
	\citenamefont {Wen}, \citenamefont {Loy}, \citenamefont {Wong},\ and\
	\citenamefont {Du}}]{Zhou2012}%
\BibitemOpen
\bibfield  {author} {\bibinfo {author} {\bibfnamefont {S.}~\bibnamefont
		{Zhou}}, \bibinfo {author} {\bibfnamefont {S.}~\bibnamefont {Zhang}},
	\bibinfo {author} {\bibfnamefont {C.}~\bibnamefont {Liu}}, \bibinfo {author}
	{\bibfnamefont {J.~F.}\ \bibnamefont {Chen}}, \bibinfo {author}
	{\bibfnamefont {J.}~\bibnamefont {Wen}}, \bibinfo {author} {\bibfnamefont
		{M.~M.~T.}\ \bibnamefont {Loy}}, \bibinfo {author} {\bibfnamefont {G.~K.~L.}\
		\bibnamefont {Wong}},\ and\ \bibinfo {author} {\bibfnamefont
		{S.}~\bibnamefont {Du}},\ }\bibfield  {title} {\bibinfo {title} {Optimal
		storage and retrieval of single-photon waveforms},\ }\href
{https://doi.org/10.1364/oe.20.024124} {\bibfield  {journal} {\bibinfo
		{journal} {Optics Express}\ }\textbf {\bibinfo {volume} {20}},\ \bibinfo
	{pages} {24124} (\bibinfo {year} {2012})}\BibitemShut {NoStop}%
\bibitem [{\citenamefont {Wang}\ \emph {et~al.}(2019)\citenamefont {Wang},
	\citenamefont {Li}, \citenamefont {Zhang}, \citenamefont {Su}, \citenamefont
	{Zhou}, \citenamefont {Liao}, \citenamefont {Du}, \citenamefont {Yan},\ and\
	\citenamefont {Zhu}}]{Wang2019}%
\BibitemOpen
\bibfield  {author} {\bibinfo {author} {\bibfnamefont {Y.}~\bibnamefont
		{Wang}}, \bibinfo {author} {\bibfnamefont {J.}~\bibnamefont {Li}}, \bibinfo
	{author} {\bibfnamefont {S.}~\bibnamefont {Zhang}}, \bibinfo {author}
	{\bibfnamefont {K.}~\bibnamefont {Su}}, \bibinfo {author} {\bibfnamefont
		{Y.}~\bibnamefont {Zhou}}, \bibinfo {author} {\bibfnamefont {K.}~\bibnamefont
		{Liao}}, \bibinfo {author} {\bibfnamefont {S.}~\bibnamefont {Du}}, \bibinfo
	{author} {\bibfnamefont {H.}~\bibnamefont {Yan}},\ and\ \bibinfo {author}
	{\bibfnamefont {S.-L.}\ \bibnamefont {Zhu}},\ }\bibfield  {title} {\bibinfo
	{title} {Efficient quantum memory for single-photon polarization qubits},\
}\href {https://doi.org/10.1038/s41566-019-0368-8} {\bibfield  {journal}
	{\bibinfo  {journal} {Nature Photonics}\ }\textbf {\bibinfo {volume} {13}},\
	\bibinfo {pages} {346} (\bibinfo {year} {2019})}\BibitemShut {NoStop}%
\bibitem [{\citenamefont {Kaczmarek}\ \emph {et~al.}(2018)\citenamefont
	{Kaczmarek}, \citenamefont {Ledingham}, \citenamefont {Brecht}, \citenamefont
	{Thomas}, \citenamefont {Thekkadath}, \citenamefont {Lazo-Arjona},
	\citenamefont {Munns}, \citenamefont {Poem}, \citenamefont {Feizpour},
	\citenamefont {Saunders}, \citenamefont {Nunn},\ and\ \citenamefont
	{Walmsley}}]{Kaczmarek2018}%
\BibitemOpen
\bibfield  {author} {\bibinfo {author} {\bibfnamefont {K.~T.}\ \bibnamefont
		{Kaczmarek}}, \bibinfo {author} {\bibfnamefont {P.~M.}\ \bibnamefont
		{Ledingham}}, \bibinfo {author} {\bibfnamefont {B.}~\bibnamefont {Brecht}},
	\bibinfo {author} {\bibfnamefont {S.~E.}\ \bibnamefont {Thomas}}, \bibinfo
	{author} {\bibfnamefont {G.~S.}\ \bibnamefont {Thekkadath}}, \bibinfo
	{author} {\bibfnamefont {O.}~\bibnamefont {Lazo-Arjona}}, \bibinfo {author}
	{\bibfnamefont {J.~H.~D.}\ \bibnamefont {Munns}}, \bibinfo {author}
	{\bibfnamefont {E.}~\bibnamefont {Poem}}, \bibinfo {author} {\bibfnamefont
		{A.}~\bibnamefont {Feizpour}}, \bibinfo {author} {\bibfnamefont {D.~J.}\
		\bibnamefont {Saunders}}, \bibinfo {author} {\bibfnamefont {J.}~\bibnamefont
		{Nunn}},\ and\ \bibinfo {author} {\bibfnamefont {I.~A.}\ \bibnamefont
		{Walmsley}},\ }\bibfield  {title} {\bibinfo {title} {High-speed noise-free
		optical quantum memory},\ }\href {https://doi.org/10.1103/PhysRevA.97.042316}
{\bibfield  {journal} {\bibinfo  {journal} {Physical Review A}\ }\textbf
	{\bibinfo {volume} {97}},\ \bibinfo {pages} {042316} (\bibinfo {year}
	{2018})}\BibitemShut {NoStop}%
\bibitem [{\citenamefont {Finkelstein}\ \emph {et~al.}(2018)\citenamefont
	{Finkelstein}, \citenamefont {Poem}, \citenamefont {Michel}, \citenamefont
	{Lahad},\ and\ \citenamefont {Firstenberg}}]{Finkelstein2018}%
\BibitemOpen
\bibfield  {author} {\bibinfo {author} {\bibfnamefont {R.}~\bibnamefont
		{Finkelstein}}, \bibinfo {author} {\bibfnamefont {E.}~\bibnamefont {Poem}},
	\bibinfo {author} {\bibfnamefont {O.}~\bibnamefont {Michel}}, \bibinfo
	{author} {\bibfnamefont {O.}~\bibnamefont {Lahad}},\ and\ \bibinfo {author}
	{\bibfnamefont {O.}~\bibnamefont {Firstenberg}},\ }\bibfield  {title}
{\bibinfo {title} {Fast, noise-free memory for photon synchronization at room
		temperature},\ }\href {https://doi.org/10.1126/sciadv.aap8598} {\bibfield
	{journal} {\bibinfo  {journal} {Science Advances}\ }\textbf {\bibinfo
		{volume} {4}},\ \bibinfo {pages} {eaap8598} (\bibinfo {year}
	{2018})}\BibitemShut {NoStop}%
\bibitem [{\citenamefont {Wolters}\ \emph {et~al.}(2017)\citenamefont
	{Wolters}, \citenamefont {Buser}, \citenamefont {Horsley}, \citenamefont
	{B{\'{e}}guin}, \citenamefont {J\"{o}ckel}, \citenamefont {Jahn},
	\citenamefont {Warburton},\ and\ \citenamefont {Treutlein}}]{Wolters2017}%
\BibitemOpen
\bibfield  {author} {\bibinfo {author} {\bibfnamefont {J.}~\bibnamefont
		{Wolters}}, \bibinfo {author} {\bibfnamefont {G.}~\bibnamefont {Buser}},
	\bibinfo {author} {\bibfnamefont {A.}~\bibnamefont {Horsley}}, \bibinfo
	{author} {\bibfnamefont {L.}~\bibnamefont {B{\'{e}}guin}}, \bibinfo {author}
	{\bibfnamefont {A.}~\bibnamefont {J\"{o}ckel}}, \bibinfo {author}
	{\bibfnamefont {J.-P.}\ \bibnamefont {Jahn}}, \bibinfo {author}
	{\bibfnamefont {R.~J.}\ \bibnamefont {Warburton}},\ and\ \bibinfo {author}
	{\bibfnamefont {P.}~\bibnamefont {Treutlein}},\ }\bibfield  {title} {\bibinfo
	{title} {Simple atomic quantum memory suitable for semiconductor quantum dot
		single photons},\ }\href {https://doi.org/10.1103/physrevlett.119.060502}
{\bibfield  {journal} {\bibinfo  {journal} {Physical Review Letters}\
	}\textbf {\bibinfo {volume} {119}},\ \bibinfo {pages} {060502} (\bibinfo
	{year} {2017})}\BibitemShut {NoStop}%
\bibitem [{\citenamefont {Thomas}\ \emph {et~al.}(2019)\citenamefont {Thomas},
	\citenamefont {Hird}, \citenamefont {Munns}, \citenamefont {Brecht},
	\citenamefont {Saunders}, \citenamefont {Nunn}, \citenamefont {Walmsley},\
	and\ \citenamefont {Ledingham}}]{Thomas2019}%
\BibitemOpen
\bibfield  {author} {\bibinfo {author} {\bibfnamefont {S.~E.}\ \bibnamefont
		{Thomas}}, \bibinfo {author} {\bibfnamefont {T.~M.}\ \bibnamefont {Hird}},
	\bibinfo {author} {\bibfnamefont {J.~H.~D.}\ \bibnamefont {Munns}}, \bibinfo
	{author} {\bibfnamefont {B.}~\bibnamefont {Brecht}}, \bibinfo {author}
	{\bibfnamefont {D.~J.}\ \bibnamefont {Saunders}}, \bibinfo {author}
	{\bibfnamefont {J.}~\bibnamefont {Nunn}}, \bibinfo {author} {\bibfnamefont
		{I.~A.}\ \bibnamefont {Walmsley}},\ and\ \bibinfo {author} {\bibfnamefont
		{P.~M.}\ \bibnamefont {Ledingham}},\ }\bibfield  {title} {\bibinfo {title}
	{Raman quantum memory with built-in suppression of four-wave-mixing noise},\
}\href {https://doi.org/10.1103/physreva.100.033801} {\bibfield  {journal}
	{\bibinfo  {journal} {Physical Review A}\ }\textbf {\bibinfo {volume}
		{100}},\ \bibinfo {pages} {033801} (\bibinfo {year} {2019})}\BibitemShut
{NoStop}%
\bibitem [{\citenamefont {Fleischhauer}\ and\ \citenamefont
	{Lukin}(2002)}]{Fleischhauer2002}%
\BibitemOpen
\bibfield  {author} {\bibinfo {author} {\bibfnamefont {M.}~\bibnamefont
		{Fleischhauer}}\ and\ \bibinfo {author} {\bibfnamefont {M.~D.}\ \bibnamefont
		{Lukin}},\ }\bibfield  {title} {\bibinfo {title} {Quantum memory for photons:
		Dark-state polaritons},\ }\href {https://doi.org/10.1103/physreva.65.022314}
{\bibfield  {journal} {\bibinfo  {journal} {Physical Review A}\ }\textbf
	{\bibinfo {volume} {65}},\ \bibinfo {pages} {022314} (\bibinfo {year}
	{2002})}\BibitemShut {NoStop}%
\bibitem [{\citenamefont {Yan}\ \emph {et~al.}(2001)\citenamefont {Yan},
	\citenamefont {Rickey},\ and\ \citenamefont {Zhu}}]{Yan2001}%
\BibitemOpen
\bibfield  {author} {\bibinfo {author} {\bibfnamefont {M.}~\bibnamefont
		{Yan}}, \bibinfo {author} {\bibfnamefont {E.~G.}\ \bibnamefont {Rickey}},\
	and\ \bibinfo {author} {\bibfnamefont {Y.}~\bibnamefont {Zhu}},\ }\bibfield
{title} {\bibinfo {title} {Electromagnetically induced transparency in cold
		rubidium atoms},\ }\href {https://doi.org/10.1364/josab.18.001057} {\bibfield
	{journal} {\bibinfo  {journal} {Journal of the Optical Society of America
			B}\ }\textbf {\bibinfo {volume} {18}},\ \bibinfo {pages} {1057} (\bibinfo
	{year} {2001})}\BibitemShut {NoStop}%
\bibitem [{\citenamefont {Walther}\ \emph {et~al.}(2007)\citenamefont
	{Walther}, \citenamefont {Eisaman}, \citenamefont {Andr{\'{e}}},
	\citenamefont {Massou}, \citenamefont {Fleischhauer}, \citenamefont
	{Zibrov},\ and\ \citenamefont {Lukin}}]{Walther2007}%
\BibitemOpen
\bibfield  {author} {\bibinfo {author} {\bibfnamefont {P.}~\bibnamefont
		{Walther}}, \bibinfo {author} {\bibfnamefont {M.~D.}\ \bibnamefont
		{Eisaman}}, \bibinfo {author} {\bibfnamefont {A.}~\bibnamefont
		{Andr{\'{e}}}}, \bibinfo {author} {\bibfnamefont {F.}~\bibnamefont {Massou}},
	\bibinfo {author} {\bibfnamefont {M.}~\bibnamefont {Fleischhauer}}, \bibinfo
	{author} {\bibfnamefont {A.~S.}\ \bibnamefont {Zibrov}},\ and\ \bibinfo
	{author} {\bibfnamefont {M.~D.}\ \bibnamefont {Lukin}},\ }\bibfield  {title}
{\bibinfo {title} {Generation of narrow-bandwidth single photons using
		electromagnetically induced transparency in atomic ensembles},\ }\href
{https://doi.org/10.1142/s0219749907002773} {\bibfield  {journal} {\bibinfo
		{journal} {International Journal of Quantum Information}\ }\textbf {\bibinfo
		{volume} {05}},\ \bibinfo {pages} {51} (\bibinfo {year} {2007})}\BibitemShut
{NoStop}%
\bibitem [{\citenamefont {Vurgaftman}\ and\ \citenamefont
	{Bashkansky}(2013)}]{Vurgaftman2013}%
\BibitemOpen
\bibfield  {author} {\bibinfo {author} {\bibfnamefont {I.}~\bibnamefont
		{Vurgaftman}}\ and\ \bibinfo {author} {\bibfnamefont {M.}~\bibnamefont
		{Bashkansky}},\ }\bibfield  {title} {\bibinfo {title} {Suppressing four-wave
		mixing in warm-atomic-vapor quantum memory},\ }\href
{https://doi.org/10.1103/physreva.87.063836} {\bibfield  {journal} {\bibinfo
		{journal} {Physical Review A}\ }\textbf {\bibinfo {volume} {87}},\ \bibinfo
	{pages} {063836} (\bibinfo {year} {2013})}\BibitemShut {NoStop}%
\bibitem [{\citenamefont {Rosenberry}\ \emph {et~al.}(2007)\citenamefont
	{Rosenberry}, \citenamefont {Reyes}, \citenamefont {Tupa},\ and\
	\citenamefont {Gay}}]{Rosenberry2007}%
\BibitemOpen
\bibfield  {author} {\bibinfo {author} {\bibfnamefont {M.~A.}\ \bibnamefont
		{Rosenberry}}, \bibinfo {author} {\bibfnamefont {J.~P.}\ \bibnamefont
		{Reyes}}, \bibinfo {author} {\bibfnamefont {D.}~\bibnamefont {Tupa}},\ and\
	\bibinfo {author} {\bibfnamefont {T.~J.}\ \bibnamefont {Gay}},\ }\bibfield
{title} {\bibinfo {title} {Radiation trapping in rubidium optical pumping at
		low buffer-gas pressures},\ }\href
{https://doi.org/10.1103/physreva.75.023401} {\bibfield  {journal} {\bibinfo
		{journal} {Physical Review A}\ }\textbf {\bibinfo {volume} {75}},\ \bibinfo
	{pages} {023401} (\bibinfo {year} {2007})}\BibitemShut {NoStop}%
\bibitem [{\citenamefont {Slattery}\ \emph {et~al.}(2015)\citenamefont
	{Slattery}, \citenamefont {Ma}, \citenamefont {Kuo},\ and\ \citenamefont
	{Tang}}]{Slattery2015}%
\BibitemOpen
\bibfield  {author} {\bibinfo {author} {\bibfnamefont {O.}~\bibnamefont
		{Slattery}}, \bibinfo {author} {\bibfnamefont {L.}~\bibnamefont {Ma}},
	\bibinfo {author} {\bibfnamefont {P.}~\bibnamefont {Kuo}},\ and\ \bibinfo
	{author} {\bibfnamefont {X.}~\bibnamefont {Tang}},\ }\bibfield  {title}
{\bibinfo {title} {Narrow-linewidth source of greatly non-degenerate photon
		pairs for quantum repeaters from a short singly resonant cavity},\ }\href
{https://doi.org/10.1007/s00340-015-6198-6} {\bibfield  {journal} {\bibinfo
		{journal} {Applied Physics B}\ }\textbf {\bibinfo {volume} {121}},\ \bibinfo
	{pages} {413} (\bibinfo {year} {2015})}\BibitemShut {NoStop}%
\bibitem [{\citenamefont {Ahlrichs}\ and\ \citenamefont
	{Benson}(2016)}]{Ahlrichs2016}%
\BibitemOpen
\bibfield  {author} {\bibinfo {author} {\bibfnamefont {A.}~\bibnamefont
		{Ahlrichs}}\ and\ \bibinfo {author} {\bibfnamefont {O.}~\bibnamefont
		{Benson}},\ }\bibfield  {title} {\bibinfo {title} {Bright source of
		indistinguishable photons based on cavity-enhanced parametric down-conversion
		utilizing the cluster effect},\ }\href {https://doi.org/10.1063/1.4939925}
{\bibfield  {journal} {\bibinfo  {journal} {Applied Physics Letters}\
	}\textbf {\bibinfo {volume} {108}},\ \bibinfo {pages} {021111} (\bibinfo
	{year} {2016})}\BibitemShut {NoStop}%
\bibitem [{\citenamefont {Tsai}\ and\ \citenamefont {Chen}(2018)}]{Tsai2018}%
\BibitemOpen
\bibfield  {author} {\bibinfo {author} {\bibfnamefont {P.-J.}\ \bibnamefont
		{Tsai}}\ and\ \bibinfo {author} {\bibfnamefont {Y.-C.}\ \bibnamefont
		{Chen}},\ }\bibfield  {title} {\bibinfo {title} {Ultrabright, narrow-band
		photon-pair source for atomic quantum memories},\ }\href
{https://doi.org/10.1088/2058-9565/aa86e7} {\bibfield  {journal} {\bibinfo
		{journal} {Quantum Science and Technology}\ }\textbf {\bibinfo {volume}
		{3}},\ \bibinfo {pages} {034005} (\bibinfo {year} {2018})}\BibitemShut
{NoStop}%
\bibitem [{\citenamefont {Cotting}(2021)}]{Cotting2021}%
\BibitemOpen
\bibfield  {author} {\bibinfo {author} {\bibfnamefont {B.}~\bibnamefont
		{Cotting}},\ }\emph {\bibinfo {title} {Spontaneous parametric down-conversion
		heralded single-photon source for quantum memory applications}},\ \href@noop
{} {Master's thesis},\ \bibinfo  {school} {University of Basel, \'{E}cole
	polytechnique f\'{e}d\'{e}rale de Lausanne} (\bibinfo {year}
{2021})\BibitemShut {NoStop}%
\bibitem [{\citenamefont {Jobez}\ \emph {et~al.}(2015)\citenamefont {Jobez},
	\citenamefont {Laplane}, \citenamefont {Timoney}, \citenamefont {Gisin},
	\citenamefont {Ferrier}, \citenamefont {Goldner},\ and\ \citenamefont
	{Afzelius}}]{Jobez2015}%
\BibitemOpen
\bibfield  {author} {\bibinfo {author} {\bibfnamefont {P.}~\bibnamefont
		{Jobez}}, \bibinfo {author} {\bibfnamefont {C.}~\bibnamefont {Laplane}},
	\bibinfo {author} {\bibfnamefont {N.}~\bibnamefont {Timoney}}, \bibinfo
	{author} {\bibfnamefont {N.}~\bibnamefont {Gisin}}, \bibinfo {author}
	{\bibfnamefont {A.}~\bibnamefont {Ferrier}}, \bibinfo {author} {\bibfnamefont
		{P.}~\bibnamefont {Goldner}},\ and\ \bibinfo {author} {\bibfnamefont
		{M.}~\bibnamefont {Afzelius}},\ }\bibfield  {title} {\bibinfo {title}
	{Coherent spin control at the quantum level in an ensemble-based optical
		memory},\ }\href {https://doi.org/10.1103/physrevlett.114.230502} {\bibfield
	{journal} {\bibinfo  {journal} {Physical Review Letters}\ }\textbf {\bibinfo
		{volume} {114}},\ \bibinfo {pages} {230502} (\bibinfo {year}
	{2015})}\BibitemShut {NoStop}%
\bibitem [{\citenamefont {Rakher}\ \emph {et~al.}(2013)\citenamefont {Rakher},
	\citenamefont {Warburton},\ and\ \citenamefont {Treutlein}}]{Rakher2013}%
\BibitemOpen
\bibfield  {author} {\bibinfo {author} {\bibfnamefont {M.~T.}\ \bibnamefont
		{Rakher}}, \bibinfo {author} {\bibfnamefont {R.~J.}\ \bibnamefont
		{Warburton}},\ and\ \bibinfo {author} {\bibfnamefont {P.}~\bibnamefont
		{Treutlein}},\ }\bibfield  {title} {\bibinfo {title} {Prospects for storage
		and retrieval of a quantum-dot single photon in an ultracold $^{87}${R}b
		ensemble},\ }\href {https://doi.org/10.1103/physreva.88.053834} {\bibfield
	{journal} {\bibinfo  {journal} {Physical Review A}\ }\textbf {\bibinfo
		{volume} {88}},\ \bibinfo {pages} {053834} (\bibinfo {year}
	{2013})}\BibitemShut {NoStop}%
\bibitem [{\citenamefont {Nunn}(2008)}]{Nunn2008}%
\BibitemOpen
\bibfield  {author} {\bibinfo {author} {\bibfnamefont {J.}~\bibnamefont
		{Nunn}},\ }\emph {\bibinfo {title} {Quantum memory in atomic ensembles}},\
\href
{https://www2.physics.ox.ac.uk/sites/default/files/2013-11-08/nunn_2008_pdf_10369.pdf}
{Ph.D. thesis},\ \bibinfo  {school} {St. John's College, University of
	Oxford} (\bibinfo {year} {2008})\BibitemShut {NoStop}%
\bibitem [{\citenamefont {Gorshkov}\ \emph
	{et~al.}(2007{\natexlab{a}})\citenamefont {Gorshkov}, \citenamefont
	{Andr\'{e}}, \citenamefont {Lukin},\ and\ \citenamefont
	{S{\o}rensen}}]{Gorshkov2007a}%
\BibitemOpen
\bibfield  {author} {\bibinfo {author} {\bibfnamefont {A.~V.}\ \bibnamefont
		{Gorshkov}}, \bibinfo {author} {\bibfnamefont {A.}~\bibnamefont {Andr\'{e}}},
	\bibinfo {author} {\bibfnamefont {M.~D.}\ \bibnamefont {Lukin}},\ and\
	\bibinfo {author} {\bibfnamefont {A.~S.}\ \bibnamefont {S{\o}rensen}},\
}\bibfield  {title} {\bibinfo {title} {Photon storage in {$\Lambda$}-type
		optically dense atomic media. {III}. {E}ffects of inhomogeneous broadening},\
}\href {https://doi.org/10.1103/PhysRevA.76.033806} {\bibfield  {journal}
	{\bibinfo  {journal} {Physical Review A}\ }\textbf {\bibinfo {volume} {76}},\
	\bibinfo {pages} {033806} (\bibinfo {year} {2007}{\natexlab{a}})}\BibitemShut
{NoStop}%
\bibitem [{\citenamefont {Gorshkov}\ \emph
	{et~al.}(2007{\natexlab{b}})\citenamefont {Gorshkov}, \citenamefont
	{Andr{\'{e}}}, \citenamefont {Fleischhauer}, \citenamefont {S{\o}rensen},\
	and\ \citenamefont {Lukin}}]{Gorshkov2007b}%
\BibitemOpen
\bibfield  {author} {\bibinfo {author} {\bibfnamefont {A.~V.}\ \bibnamefont
		{Gorshkov}}, \bibinfo {author} {\bibfnamefont {A.}~\bibnamefont
		{Andr{\'{e}}}}, \bibinfo {author} {\bibfnamefont {M.}~\bibnamefont
		{Fleischhauer}}, \bibinfo {author} {\bibfnamefont {A.~S.}\ \bibnamefont
		{S{\o}rensen}},\ and\ \bibinfo {author} {\bibfnamefont {M.~D.}\ \bibnamefont
		{Lukin}},\ }\bibfield  {title} {\bibinfo {title} {Universal approach to
		optimal photon storage in atomic media},\ }\href
{https://doi.org/10.1103/physrevlett.98.123601} {\bibfield  {journal}
	{\bibinfo  {journal} {Physical Review Letters}\ }\textbf {\bibinfo {volume}
		{98}},\ \bibinfo {pages} {123601} (\bibinfo {year}
	{2007}{\natexlab{b}})}\BibitemShut {NoStop}%
\bibitem [{\citenamefont {Gorshkov}\ \emph {et~al.}(2008)\citenamefont
	{Gorshkov}, \citenamefont {Calarco}, \citenamefont {Lukin},\ and\
	\citenamefont {S\o{}rensen}}]{Gorshkov2008}%
\BibitemOpen
\bibfield  {author} {\bibinfo {author} {\bibfnamefont {A.~V.}\ \bibnamefont
		{Gorshkov}}, \bibinfo {author} {\bibfnamefont {T.}~\bibnamefont {Calarco}},
	\bibinfo {author} {\bibfnamefont {M.~D.}\ \bibnamefont {Lukin}},\ and\
	\bibinfo {author} {\bibfnamefont {A.~S.}\ \bibnamefont {S\o{}rensen}},\
}\bibfield  {title} {\bibinfo {title} {Photon storage in
		$\ensuremath{\Lambda}$-type optically dense atomic media. {IV}. optimal
		control using gradient ascent},\ }\href
{https://doi.org/10.1103/PhysRevA.77.043806} {\bibfield  {journal} {\bibinfo
		{journal} {Phys. Rev. A}\ }\textbf {\bibinfo {volume} {77}},\ \bibinfo
	{pages} {043806} (\bibinfo {year} {2008})}\BibitemShut {NoStop}%
\bibitem [{\citenamefont {Phillips}\ \emph {et~al.}(2008)\citenamefont
	{Phillips}, \citenamefont {Gorshkov},\ and\ \citenamefont
	{Novikova}}]{Phillips2008}%
\BibitemOpen
\bibfield  {author} {\bibinfo {author} {\bibfnamefont {N.~B.}\ \bibnamefont
		{Phillips}}, \bibinfo {author} {\bibfnamefont {A.~V.}\ \bibnamefont
		{Gorshkov}},\ and\ \bibinfo {author} {\bibfnamefont {I.}~\bibnamefont
		{Novikova}},\ }\bibfield  {title} {\bibinfo {title} {Optimal light storage in
		atomic vapor},\ }\href {https://doi.org/10.1103/physreva.78.023801}
{\bibfield  {journal} {\bibinfo  {journal} {Physical Review A}\ }\textbf
	{\bibinfo {volume} {78}},\ \bibinfo {pages} {023801} (\bibinfo {year}
	{2008})}\BibitemShut {NoStop}%
\bibitem [{\citenamefont {Hong}\ \emph {et~al.}(1987)\citenamefont {Hong},
	\citenamefont {Ou},\ and\ \citenamefont {Mandel}}]{Hong1987}%
\BibitemOpen
\bibfield  {author} {\bibinfo {author} {\bibfnamefont {C.~K.}\ \bibnamefont
		{Hong}}, \bibinfo {author} {\bibfnamefont {Z.~Y.}\ \bibnamefont {Ou}},\ and\
	\bibinfo {author} {\bibfnamefont {L.}~\bibnamefont {Mandel}},\ }\bibfield
{title} {\bibinfo {title} {Measurement of subpicosecond time intervals
		between two photons by interference},\ }\href
{https://doi.org/10.1103/physrevlett.59.2044} {\bibfield  {journal} {\bibinfo
		{journal} {Physical Review Letters}\ }\textbf {\bibinfo {volume} {59}},\
	\bibinfo {pages} {2044} (\bibinfo {year} {1987})}\BibitemShut {NoStop}%
\bibitem [{\citenamefont {M{\o}ller}\ \emph {et~al.}(2017)\citenamefont
	{M{\o}ller}, \citenamefont {Thomas}, \citenamefont {Vasilakis}, \citenamefont
	{Zeuthen}, \citenamefont {Tsaturyan}, \citenamefont {Balabas}, \citenamefont
	{Jensen}, \citenamefont {Schliesser}, \citenamefont {Hammerer},\ and\
	\citenamefont {Polzik}}]{Moeller2017}%
\BibitemOpen
\bibfield  {author} {\bibinfo {author} {\bibfnamefont {C.~B.}\ \bibnamefont
		{M{\o}ller}}, \bibinfo {author} {\bibfnamefont {R.~A.}\ \bibnamefont
		{Thomas}}, \bibinfo {author} {\bibfnamefont {G.}~\bibnamefont {Vasilakis}},
	\bibinfo {author} {\bibfnamefont {E.}~\bibnamefont {Zeuthen}}, \bibinfo
	{author} {\bibfnamefont {Y.}~\bibnamefont {Tsaturyan}}, \bibinfo {author}
	{\bibfnamefont {M.}~\bibnamefont {Balabas}}, \bibinfo {author} {\bibfnamefont
		{K.}~\bibnamefont {Jensen}}, \bibinfo {author} {\bibfnamefont
		{A.}~\bibnamefont {Schliesser}}, \bibinfo {author} {\bibfnamefont
		{K.}~\bibnamefont {Hammerer}},\ and\ \bibinfo {author} {\bibfnamefont
		{E.~S.}\ \bibnamefont {Polzik}},\ }\bibfield  {title} {\bibinfo {title}
	{Quantum back-action-evading measurement of motion in a negative mass
		reference frame},\ }\href {https://doi.org/10.1038/nature22980} {\bibfield
	{journal} {\bibinfo  {journal} {Nature}\ }\textbf {\bibinfo {volume} {547}},\
	\bibinfo {pages} {191} (\bibinfo {year} {2017})}\BibitemShut {NoStop}%
\bibitem [{\citenamefont {Corsini}\ \emph {et~al.}(2013)\citenamefont
	{Corsini}, \citenamefont {Karaulanov}, \citenamefont {Balabas},\ and\
	\citenamefont {Budker}}]{Corsini2013}%
\BibitemOpen
\bibfield  {author} {\bibinfo {author} {\bibfnamefont {E.~P.}\ \bibnamefont
		{Corsini}}, \bibinfo {author} {\bibfnamefont {T.}~\bibnamefont {Karaulanov}},
	\bibinfo {author} {\bibfnamefont {M.}~\bibnamefont {Balabas}},\ and\ \bibinfo
	{author} {\bibfnamefont {D.}~\bibnamefont {Budker}},\ }\bibfield  {title}
{\bibinfo {title} {Hyperfine frequency shift and {Z}eeman relaxation in
		alkali-metal-vapor cells with antirelaxation alkene coating},\ }\href
{https://doi.org/10.1103/physreva.87.022901} {\bibfield  {journal} {\bibinfo
		{journal} {Physical Review A}\ }\textbf {\bibinfo {volume} {87}},\ \bibinfo
	{pages} {022901} (\bibinfo {year} {2013})}\BibitemShut {NoStop}%
	
\end{thebibliography}

\end{document}